\documentclass{iopjournal}
\usepackage[T1]{fontenc}
\usepackage{lmodern}

\usepackage{graphicx}
\usepackage{dcolumn}
\usepackage{comment}
\usepackage{bm}
\usepackage{xcolor}
\usepackage{ragged2e}
\definecolor{darkblue}{rgb}{0,0,0.6}
\definecolor{forestgreen}{HTML}{228B22}
\colorlet{green}{forestgreen}
\usepackage[normalem]{ulem}
\usepackage{hyperref}
\usepackage{placeins}
\hypersetup{colorlinks=true, linkcolor=darkblue, citecolor=darkblue, urlcolor=darkblue}

\usepackage{amsmath,amssymb}

\usepackage[
  backend=biber,
  style=numeric,
  citestyle=numeric-comp,
  sorting=none,
  maxbibnames=999,
  doi=false,url=false,isbn=false,eprint=false
]{biblatex}

\DeclareFieldFormat{journaltitle}{%
  \mkbibemph{%
    \ifboolexpr{ test {\iffieldundef{doi}} and test {\iffieldundef{url}} }
      {#1}
      {\href{\iffieldundef{doi}{\thefield{url}}{https://doi.org/\thefield{doi}}}{#1}}%
  }%
}

\DeclareFieldFormat{pages}{#1}
\renewbibmacro{in:}{}
\AtEveryBibitem{\clearfield{month}\clearfield{day}}
\AtBeginDocument{\justifying}

\DeclareListFormat{publisher}{%
  \ifboolexpr{ test {\iffieldundef{doi}} and test {\iffieldundef{url}} }
    {#1}
    {\href{\iffieldundef{doi}{\thefield{url}}{https://doi.org/\thefield{doi}}}{#1}}%
  \ifnum\value{listcount}<\value{liststop}\addcomma\space\fi
}

\DeclareCiteCommand{\onlinecite}
  {\ifnumequal{\value{citetotal}}{2}
     {}
     {}%
   \usebibmacro{cite:init}%
   \usebibmacro{prenote}}
  {\usebibmacro{citeindex}%
   \usebibmacro{cite:comp}}
  {}
  {\usebibmacro{cite:dump}%
   \usebibmacro{postnote}}

\usepackage{etoolbox} 
\makeatletter
\patchcmd{\@makecaption}{\raggedright}{\justifying}{}{}
\makeatother

\definecolor{lightgray}{gray}{0.8}

\begin{document}

\articletype{Paper}

\title{Metastability and nucleation of bubbles in active chiral systems}

\author{Raphaël Maire$^{1, *}$\orcid{0009-0007-7530-9626}, Alessandro Petrini$^2$\orcid{0009-0006-7751-7210}, Umberto Marini Bettolo Marconi$^3$\orcid{0000-0002-2764-8259} and Lorenzo Caprini$^{2,*}$\orcid{0000-0003-1384-0917}}

\affil{$^1$Department of Condensed Matter, University of Barcelona, 08028 Barcelona, Spain}

\affil{$^2$Sapienza University of Rome, IT-00185 Rome, Italy}

\affil{$^3$Universit\`a di Camerino, Scuola di Scienze e Tecnologie, 62032, Camerino, Italy}

\affil{$^*$Authors to whom any correspondence should be addressed.}

\email{maire@ub.edu}
\email{lorenzo.caprini@uniroma1.it}

\keywords{active matter, chiral active fluids, odd viscosity, bubble nucleation, metastability, effective potential, collective-coordinate reduction, hydrodynamic instabilities, event-driven molecular dynamics}

\begin{abstract}
Bubbles frequently emerge in dense chiral active fluids, yet it remains unclear whether they result from an instability of the homogeneous state or from rare nucleation events within a metastable phase. Here we develop a nucleation theory for chiral bubble formation by reducing the hydrodynamic equations to a single collective coordinate, the bubble radius. This reduction yields an effective nonlinear potential that captures the radial dynamics and predicts distinct stable, metastable, and unstable regimes of the homogeneous state. In the metastable regime, the theory identifies a critical bubble radius and an associated effective mechanical-work barrier, while its large-radius behavior determines whether bubble growth saturates at a finite size or becomes unbounded. We further analyze the stability of circular bubbles against noncircular perturbations, finding growing modes that can lead to elongation and breakup. The theoretical predictions are tested against event-driven molecular-dynamics simulations of chiral hard disks, which reproduce the predicted trends in the critical radius, nucleation barrier, and selected bubble size. Our results establish a unified framework for stability and nucleation in chiral active fluids and suggest routes to control bubble formation in experiments with spinning colloids and chiral spinners.

\end{abstract}

\section{Introduction}

Metastability, introduced by Ostwald~\cite{ostwald1891lehrbuch} in the late nineteenth century and later formalized by Gibbs~\cite{gibbs1960elementary}, refers to the tendency of a system to persist for long times in a locally stable state before rare fluctuations induce a transition to another state. In equilibrium statistical mechanics, this picture is naturally formulated in terms of a free-energy landscape parameterized by one or more collective coordinates. Stable and metastable states correspond to local minima of the free energy, separated from competing states by barriers that can only be crossed through sufficiently large fluctuations. The resulting kinetics, characterized by rare events punctuated by abrupt transformations, underlies a broad range of phenomena, including cavitation in liquids~\cite{caupin2006cavitation}, phase transitions from supersaturated, supercooled, or superheated states~\cite{onuki2002phase}, magnetization reversal~\cite{hubert2008magnetic}, and chemical or biochemical assembly~\cite{whitesides2002self}. In this picture, the transition proceeds through a critical configuration, or nucleus, located near the top of the relevant free-energy barrier. Classical reaction-rate theories due to Kramers, Eyring, Langer, and others relate the transition rate to this barrier height, leading to the characteristic Arrhenius suppression of rare events~\cite{hanggi1990reaction}.

This equilibrium picture, however, does not directly carry over to systems driven far from equilibrium. For stochastic dynamics, rare transitions can instead be formulated in terms of the probability of entire trajectories, which in the weak-noise limit is described by a large-deviation action~\cite{Freidlin_Wentzell_1984,touchette2009large,zakine2023minimum}. Metastable states can still emerge as long-lived states of the dynamics, but the relevant landscape is no longer a thermodynamic free energy, and transition pathways can differ qualitatively from their equilibrium counterparts. In particular, broken time-reversal symmetry can make forward and backward transition paths distinct~\cite{yao2025interfacial,schuttler2025nonreciprocal,zakine2024unveiling}, while nonequilibrium contributions can also modify the transition rates beyond their leading exponential dependence~\cite{bouchet2016generalisation}. These considerations are especially relevant to active matter~\cite{marchetti2013hydrodynamics,elgeti2015physics,bechinger2016active}, whose constituents continuously consume energy to generate persistent motion and mechanical stresses. Recent work has demonstrated that metastability and nucleation can arise in active systems, including flocking models and self-propelled particle systems~\cite{besse2022metastability,benvegnen2023metastability,levis2017active,PhysRevLett.114.068101,richard2016nucleation,redner2016classical,omar2021phase}. Classical nucleation theory has been extended to active phase separation~\cite{cates2023classical,langford2024mechanics} and, more recently, to nonequilibrium systems with nonconserved order parameters, including models of active matter and population dynamics~\cite{chatzittofi2026nonequilibrium,ziethen2026nucleation}. In hyperuniform fluids, suppressed large-scale fluctuations can further modify the nucleation barrier and its dependence on nucleus size~\cite{maire2026hyperuniformnucleation}. Active phase coexistence can also exhibit equilibrium-like features such as binodals, spinodals, and generalized Maxwell constructions~\cite{solon2018generalized,burekovic2026active}, although these analogies may break down in genuinely nonequilibrium or non-reciprocal settings~\cite{omar2023mechanical,mason2025dynamical}.

Within this broader class, chiral active matter \cite{lowen2016chirality, liebchen2022chiral, caprini2026modeling} provides a particularly interesting setting in which to study metastability and nucleation. Active systems that break parity symmetry are called chiral and can exhibit persistent rotational or circular motion due to intrinsic handedness in their propulsion mechanism and body shape~\cite{lowen2016chirality} or microscopic interactions~\cite{caprini2026modeling}. Chiral dynamics can generate collective phenomena \cite{liebchen2017collective,huang2020dynamical,liao2018clustering,zhang2020reconfigurable,kreienkamp2022clustering,lei2023collective,jeong2025crystallization,musacchio2026circling} and transport coefficients absent from ordinary parity- and time-reversal-invariant fluids, including antisymmetric and more general parity-odd contributions~\cite{fruchart2023odd}, ranging from odd diffusivity \cite{hargus2021odd,kalz2022collisions,vega2022diffusive,kalz2024oscillatory,abdoli2026dynamical,caprini2025active} to odd viscosity \cite{avron1998odd,banerjee2017odd,markovich2021odd,mecke2023simultaneous,souslov2019topological,hosaka2021nonreciprocal,maire2026Microscopic} and elasticity \cite{scheibner2020odd,lee2025odd,braverman2021topological,ishimoto2022self,caprini2025odd,zhou2026curved}. These terms can produce transverse responses~\cite{maire2026chiralintruder,maire2026intrudertheories}, circulating flows, and unusual hydrodynamic modes~\cite{workamp2018symmetry,caprini2019active,mecke2024emergent,negi2023geometry,caprini2023chiral,caporusso2024phase,markovich2025chiral,abanov2018odd,burekovic2026bulk,kuroda2026designing,petrini2026curvature, reyes2026steadybasestatestwodimensional}, and can qualitatively alter the mechanical response of both fluids and solids~\cite{everts2024dissipative,khain2022stokes,yuan2023stokesian,ganeshan2017odd,scheibner2020odd,tan2022odd,huang2025anomalous}. Recent particle-based simulations of inertial chiral active fluids have reported the spontaneous formation of bubbles, or cavities, within an otherwise dense phase~\cite{caprini2025Bubble,shen2023collective,digregorio2025phase,petrini2026sparkling} (see Fig.~\ref{fig:overview}(a) for a representative bubble configuration). These bubbles are surrounded by persistent azimuthal flows, sustained by the interplay of pressure, dissipation, and centrifugal forces \cite{Marconi2026hydrodynamics}. Yet the observation of a bubble phase does not by itself reveal how it emerges. Does a sufficiently dense homogeneous state become linearly unstable, so that arbitrarily small fluctuations grow into a cavity? Or can the homogeneous state remain locally stable, with bubbles appearing only through rare, finite-amplitude fluctuations? If the latter occurs, what sets the critical nucleus and the associated nucleation barrier? And once a bubble has formed, is its size selected by the dynamics, or does it continue to grow indefinitely? These questions are the chiral analogue of the classical distinction between spinodal decomposition and nucleation, but cannot be answered from existing hydrodynamic or particle-based descriptions alone.

Here, we address these questions by developing a minimal theory of bubble nucleation in the bubble phase induced by odd interactions (BIO). Starting from the incompressible odd-hydrodynamic framework of Ref.~\onlinecite{Marconi2026hydrodynamics}, we retain viscous dissipation and introduce a general density-dependent torque density to represent the coarse-grained effect of microscopic transverse forces. We then project the hydrodynamic equations onto the collective coordinate given by the bubble radius, obtaining an effective nonlinear potential governing the time evolution of a bubble. This reduction allows us to distinguish stable, metastable, and unstable homogeneous states. In the metastable regime, the potential yields a critical nucleus and its associated barrier, while its large-radius behavior determines whether a bubble reaches a finite selected size or undergoes runaway growth. The single-coordinate description also has a well-defined limitation: a circular bubble need not remain circular. We therefore analyze the stability of the circular solution to noncircular perturbations, identifying conditions under which deformation can lead to elongation and breakup, as observed in simulations~\cite{shen2023collective,petrini2026sparkling}. We test these predictions using event-driven molecular-dynamics simulations of the minimal chiral hard-disk model introduced in
Ref.~\onlinecite{maire2026Microscopic}. The simulations capture the predicted trends in the critical radius, nucleation barrier, and selected bubble size, and support the theoretical distinction between stable and metastable regimes.

The paper is organized as follows. Section~\ref{sec:1} introduces the microscopic model, while Sec.~\ref{sec:overview} presents an overview of the main numerical and theoretical results of the paper.
We derive the nucleation theory, obtaining an effective dynamics for a circular cavity and connecting the mechanical work of bubble formation to the nucleation barrier in Sec.~\ref{sec:2}, while Sec.~\ref{sec:3} presents the main outcome of the theory, determining the stability of the homogeneous state.
Section~\ref{Sec:Numerics} reports numerical evidence for the metastability of the homogeneous state and compares simulations with theory, while Sec.~\ref{sec:4} focuses on the finite-size and shape instability of an isolated bubble. Finally, Sec.~\ref{sec:5} discusses limitations and future directions, while the appendices detail the mapping from microscopic to hydrodynamic parameters, the fluctuating bubble dynamics, and the analyses of radial and noncircular stability.

\section{Model}
\label{sec:1}

\begin{figure}[!ht]
    \centering
    \includegraphics[width=0.99\linewidth]{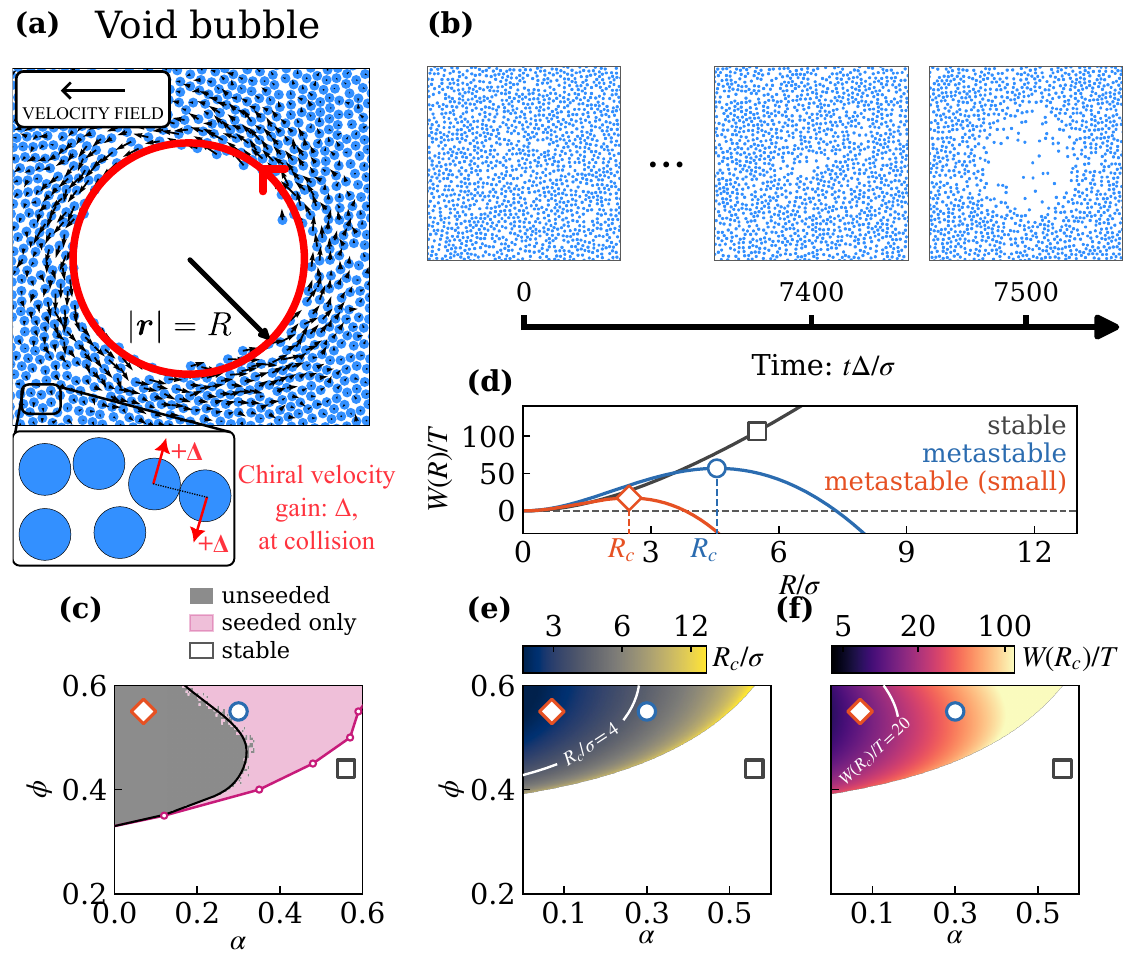}
    \caption{
    Bubble nucleation driven by chiral collisions: simulations and theory.
    (a) Snapshot of a void cavity of radius $R$ surrounded by a circulating flow. Here, black arrows denote the particle velocity, while the red arrow shows the direction of edge currents at the bubble boundary.
    (b) Typical time evolution of an initially homogeneous system leading to bubble formation. Parameters are fixed at $\alpha=0.29$ and $\phi=0.438$, with snapshots taken at $t\Delta/\sigma={0,7400,7500}$.
    (c) Phase diagram in the plane of packing fraction $\phi$ and restitution coefficient $\alpha$, where low $\alpha$ is equivalent to large chirality in dimensionless units. Colors distinguish between different simulation regimes within the observation time: i) a spontaneous bubble formation from an initially homogeneous state (gray region), termed ``unseeded'', if a bubble nucleates spontaneously within $t|\Delta|/\sigma<10^4$; ii) bubbles sustained only after seeding (pink region), termed ``seeded only'', if a preformed bubble persists despite no spontaneous nucleation within this time. iii) a stable homogeneous state (white region), termed ``stable'', if none of the previous conditions occur. To prepare a seed, we let a bubble form and begin growing at low $\alpha$, then increase $\alpha$ to the target value at fixed $\phi$ and monitor whether the bubble survives or collapses. 
    (d) Mechanical work of bubble formation $W(R)/T$ for a stable homogeneous state and two metastable states. (e)-(f) Theoretical predictions of the critical radius $R_c/\sigma$ and mechanical work of formation $W(R_c)/T$, in the plane of $\phi$ and $\alpha$. The critical radius $R_c/\sigma$ is the smaller positive root of $F(R)=0$ (Eq.~\eqref{eq:simplfiied}) and $W(R_c)/T$ is obtained by integration of $F$. A white region has no critical radius. The square, circle, and diamond identify the state points $(\alpha,\phi)=(0.56,0.44)$, $(0.30,0.55)$, and $(0.07,0.55)$, corresponding to the curves in (d). For all figures, the damping is $\gamma\sigma/|\Delta|=0.05$ while for particle simulations $N=16000$. 
    }
    \label{fig:overview}
\end{figure}

The microscopic model consists of identical disks of mass $m_p$ and diameter $\sigma$ undergoing instantaneous binary collisions. When two chiral particles, labeled $1$ and $2$, reach contact at $|\bm r_1-\bm r_2|=\sigma$, their velocities change according to the following collision rule
\begin{equation}
    \bm v_1'= \bm v_1 - \dfrac{1+\alpha}{2}(\bm v_{12}\cdot \hat{\bm\sigma}_{12})\hat{\bm\sigma}_{12} -  \Delta\hat{\bm\sigma}_{12}^\perp\,, 
    \quad \bm v_2'= \bm v_2 + \dfrac{1+\alpha}{2}(\bm v_{12}\cdot \hat{\bm\sigma}_{12})\hat{\bm\sigma}_{12} + \Delta\hat{\bm\sigma}_{12}^\perp\,,
\label{eq:collision_ruleoverview}
\end{equation}
where $\bm v_{12}=\bm v_1-\bm v_2$ corresponds to the relative velocity, $\hat{\bm\sigma}_{12}=(\bm r_1-\bm r_2)/\sigma$ is the unit vector connecting the particle centers, and $\hat{\bm\sigma}_{12}^{\perp}=\bm\varepsilon\cdot\hat{\bm\sigma}_{12}$, with $\bm\varepsilon$ the two-dimensional Levi-Civita pseudotensor ($\varepsilon_{xx}=\varepsilon_{yy}=0$ and $\varepsilon_{xy}=-\varepsilon_{yx}=1$), identifies the orthogonal direction to $\hat{\bm\sigma}_{12}$. 

As in granular media, collisions are dissipative, with restitution coefficient $\alpha<1$. Chirality is introduced by imparting opposite transverse velocity increments of magnitude $|\Delta|$ to the two particles, as sketched in Fig.~\ref{fig:overview}(a). These increments are the signature of the particle chirality, manifesting in the collision as an injection of angular momentum through transverse kicks with fixed amplitude and handedness controlled by the sign of $\Delta$.
Between collisions, the particles undergo dissipative free flight, with velocity
\begin{equation}
    \bm v(t)=\bm v(0)\exp(-\gamma t)\,,
    \label{eq:v_gamma}
\end{equation}
where $\gamma$ is the damping rate and $\bm v(0)$ is the particle velocity at the collision time. As we will see below, damping helps stabilize a bubble; however, excessive damping suppresses the circulating flow and, consequently, the centrifugal forces required for bubble formation. 
We simulate this model for $N$ colliding particles using an event-driven molecular-dynamics algorithm~\cite{smallenburg2022efficient} in a periodic square box of size $L\times L$. In our numerical study, we vary the packing fraction $\phi=N\pi\sigma^2/(4L^2)$ and the coefficient of restitution $\alpha$. Since $m_p\Delta^2$ is used as the unit of energy, $\alpha$ can be regarded as the control parameter governing the effectiveness of particle chirality~\cite{maire2026Microscopic}, i.e., the relative strength of the transverse kicks: smaller values of $\alpha$ correspond to stronger particle chirality. Accordingly, we will often use the term chirality to refer to $\alpha$.

\section{Overview of the results: stability, metastability and instability of the homogeneous phase}
\label{sec:overview}
\vskip5pt
\noindent
\textbf{Numerical evidence of bubbles and metastability --}
Simulations show that, for suitable parameters, an initially homogeneous system can develop bubbles. However, the formation of a bubble can require very long times, before which the system remains homogeneous (see the snapshots in Fig.~\ref{fig:overview}(b) at different times). To explore this dynamics, we systematically scan bubble formation as a function of packing fraction $\phi$ and chirality $\alpha$ in Fig.~\ref{fig:overview}(c), revealing three distinct regimes.
At low $\phi$ and large $\alpha$, no bubbles form, and the system remains homogeneous (white region in Fig.~\ref{fig:overview}(c)). Upon increasing $\phi$, for sufficiently low $\alpha$, bubbles spontaneously emerge from an initially homogeneous state (grey region in Fig.~\ref{fig:overview}(c)). We refer to this region as ``unseeded'' because bubble formation does not require an initial seed. 
Moreover, at larger $\alpha$ and sufficiently high $\phi$, bubbles do not form spontaneously within the simulation time, and their formation must be triggered by imposing an initial small ‘‘seed'' (pink region in Fig.~\ref{fig:overview}(c), termed ``seeded only''). This dependence on the initial condition provides additional evidence for the possible metastability of the homogeneous state to nucleation of void bubbles.

This bubble formation and stability have previously been explained qualitatively through a scaling argument, in which edge currents at the bubble boundary generate a centrifugal force capable of balancing the pressure exerted by the surrounding particles~\cite{caprini2025Bubble}. Subsequently, a hydrodynamic description derived from the microscopic equations was used to characterize the bubble phase through a linear stability analysis of the homogeneous state~\cite{Marconi2026hydrodynamics}. 
This hydrodynamic theory predicts that a chiral active fluid is described by an Ekman-Navier-Stokes equation with two additional terms compared to conventional fluids: an odd viscosity term (see Ref.~\onlinecite{fruchart2023odd} for a review) and a torque density in the form of a skew-density gradient.
This hydrodynamic description provides the starting point for a nucleation theory that distinguishes stable and metastable homogeneous states. To our knowledge, such a framework has not yet been developed for bubble formation in chiral active fluids.

\vskip5pt
\noindent
\textbf{Nucleation theory for bubbles in chiral active fluids --}
To develop a nucleation theory, we consider a circular void of radius $R$ surrounded by fluid (Fig.~\ref{fig:overview}(a)) and ask whether the resulting flow makes the cavity shrink, grow, or remain stationary.  Under hypotheses detailed in Sec.~\ref{sec:2} and Appendix~\ref{app:fluctuating_rayleigh_plesset}, the radial force balance reduces the full hydrodynamics to an effective equation for the bubble radius,
\begin{equation}
    \mathcal Z(R)\dot R=-\dfrac{\partial W}{\partial R}\,,  \qquad F(R)=-\dfrac{\partial W}{\partial R}\,,
\end{equation}
where $W(R)$ is the mechanical work required to form quasistatically a bubble from the homogeneous state, $F(R)$ is the associated radial force and $\mathcal Z(R)$ is an effective friction coefficient. Depending on the sign of $F$, the bubble expands ($F(R)>0$) or collapses ($F(R)<0$). For example, neglecting odd viscosity and line tension, we can obtain an explicit expression for $F(R)$ as a function of the model parameters:
\begin{equation}
\dfrac{F(R)}{2\pi R}=-p_\infty+\frac{\tau^2}{2\eta\gamma}
g\left(\frac{R}{\sqrt{\eta/(\gamma\rho)}}\right)\,, 
\label{eq:simplfiied}
\end{equation}
where $g(x)>0$ is a positive non-monotonic function (later given in Eq.~\eqref{eq:g}). 
Here, $\gamma$ and $\eta$ correspond to the damping rate and shear viscosity, respectively, while $\rho$ and $\tau=\tau(\rho)$ are the density and torque density contributions appearing in the hydrodynamic description (see Ref.~\onlinecite{Marconi2026hydrodynamics} or Sec.~\ref{sec:2}). These latter two quantities are spatially uniform in the incompressible exterior fluid. The first term on the right-hand side of Eq.~\eqref{eq:simplfiied} corresponds to the far-field pressure $\lim_{r\to\infty}\mathtt p(r)=p_\infty>0$, which tends to suppress the bubble because of its negative sign. By contrast, the torque density drives a circulating edge current, whose centrifugal contribution to the radial traction is proportional to $\tau^2$ and favors expansion. The competition between these two terms provides a mechanism for the formation of a bubble with a finite radius.

Figure~\ref{fig:overview}(d) illustrates this competition by plotting the mechanical work $W(R)$, obtained by integrating $W(R)=-\int_0^R dR' F(R')$. If $W(R)$ increases monotonically, then $F(R)<0$ and a cavity of any radius collapses so that a homogeneous state is recovered. If $W(R)$ instead reaches a maximum at a critical radius $R_c$, cavities smaller than $R_c$ collapse, while those slightly larger than $R_c$ grow. This picture resembles equilibrium nucleation, where fluctuations must overcome a free-energy barrier before a nucleus can grow.

\vskip5pt
\noindent
\textbf{Comparison between theory and simulations --}
Using the approximate relations between the hydrodynamic quantities and the model parameters (Eqs.~\eqref{eq:collision_ruleoverview} and~\eqref{eq:v_gamma}) reported in Appendix~\ref{app:transport}, we estimate the critical radius and mechanical work of formation for the microscopic model. Figures~\ref{fig:overview}(e) and (f) show the resulting maps of $R_c/\sigma$ and the work barrier in units of the kinetic temperature $T=\frac{m_p}{2N}\sum_{i}\bm v_i^2$. In the white regions, $W(R)$ increases monotonically and the homogeneous state is predicted to be stable against bubble formation since any bubble of finite size would shrink. In the colored regions, a bubble can grow once fluctuations carry it beyond the critical radius $R_c$ by overcoming the nucleation barrier. At high packing fraction and low $\alpha$, both the critical radius and the barrier are small because the torque-driven edge currents generate strong centrifugal forces acting against the pressure that tends to shrink the cavity. Bubble formation is therefore easier. At low packing fraction, interactions become rare, weakening the torque density $\tau$ and the resulting edge currents. Pressure, however, retains the ideal contribution proportional to the density, so the centrifugal forces are too weak to overcome it and sustain bubble growth.

We use $\exp[-W(R_c)/T]$ as an approximate activation factor for nucleation, with the kinetic temperature $T$ setting the fluctuation scale. Although in our nonequilibrium fluid the mechanical work $W$ is not the exact large-deviation function, the fluctuating theory developed below provides a basis for this qualitative interpretation and clarifies its limitations (Appendix~\ref{app:fluctuating_rayleigh_plesset}). The comparison uses $\eta_o=0$ and the approximate transport coefficients given in Appendix~\ref{app:transport}.
The predictions capture the main simulation trends in Fig.~\ref{fig:overview}(c), showing a similar phase diagram.
Indeed, both theory and simulations predict stability at low packing fraction and large $\alpha$, while the region requiring seeding in simulations corresponds to a region where $W(R_c)$ is large. The barrier contours also bend back in a manner similar to the boundary of the unseeded region. 
This picture is consistent with the long waiting time illustrated in Fig.~\ref{fig:overview}(b) and the strong parameter dependence of nucleation times observed in the simulations. However, the agreement is only qualitative since spontaneous nucleation occurs in the simulations even where we predict $W(R_c)/T\sim25$, for which the approximate activation factor would suggest very rare events.

This overview of the results interprets bubble formation as a nucleation process, but leaves several questions open. We have neglected odd viscosity, although it can strongly affect stability, and have not yet examined the dynamics beyond the nucleation barrier. In the following sections, we derive the theory in detail and investigate the subsequent growth, size selection, and shape stability of the bubble.

\section{Derivation of the nucleation theory}
\label{sec:2}

\subsection{Macroscopic model: the odd Navier-Stokes equation}

To develop our nucleation theory, we start from the hydrodynamic description of chiral fluids derived from a Langevin dynamics with odd (transverse) interactions~\cite{Marconi2026hydrodynamics} or from the microscopic model~\cite{maire2026Microscopic}, also introduced in Sec.~\ref{sec:1}.
Taking the incompressible limit of these hydrodynamic descriptions gives a two-dimensional incompressible odd Navier--Stokes equation, retaining viscous contributions and a general torque density $\tau(\rho)$:
\begin{equation}   
\label{eq:odd-navier_stokes}
   \rho (\partial_t \bm {u} + \bm {u}\cdot\bm  \nabla \bm {u})  = -\gamma\rho\bm  u-\bm  \nabla \mathtt{p}+ \bm\varepsilon\cdot \bm  \nabla\tau + \eta\bm\nabla^2\bm u + \eta_o\bm\varepsilon\cdot \bm\nabla^2\bm u\,,  \qquad \bm\nabla\cdot\bm u=0\,.
\end{equation}
where $\rho$ and $\bm u$ are the mass density and velocity field, respectively. The coefficients $\eta$ and $\eta_o$ are the shear and odd viscosities, respectively, while the incompressibility condition eliminates most of the other viscous contributions~\cite{fruchart2023odd}. The mechanical pressure $\mathtt p$ enforces this constraint as a Lagrange multiplier~\cite{landau1987fluid}, with its far-field value fixed by the equation of state\footnote{Here, an equation of state relates the homogeneous virial pressure to homogeneous hydrodynamic fields retained, such as $\rho$. Its relation to the \emph{exact} wall forces is not determined by the hydrodynamic formulation alone~\cite{solon2015notstatefunction}.}: $\mathtt p(r\to\infty)=p(\rho)\equiv p_\infty$.
Consequently, chirality is responsible not only for odd viscosity but also for a force density $\bm\varepsilon\cdot\bm\nabla\tau$ perpendicular to the gradient of the torque density $\tau$~\cite{Marconi2026hydrodynamics,maire2026Microscopic}. Here, $\tau$ is a pseudoscalar whose gradient is rotated by $90^\circ$ by the Levi-Civita symbol $\bm\varepsilon$.
This force is the divergence of the antisymmetric stress $\tau\bm\varepsilon$, just as the pressure force arises from the isotropic stress $-p\bm 1$. At a boundary, pressure acts normally, whereas torque drives tangential motion.

The microscopic origin of these two homogeneous stresses is evident when the pair force is written as the sum of central and transverse components~\cite{Marconi2026hydrodynamics} $\bm f_{ij}=f_{\parallel}(r_{ij})\hat{\bm r}_{ij}+f_{\perp}(r_{ij})(\bm\varepsilon\cdot\hat{\bm r}_{ij})\,,$ where $r_{ij}=|\bm r_i-\bm r_j|$ and $\hat{\bm r}_{ij}=(\bm r_i-\bm r_j)/r_{ij}$. 
Transverse force components typically emerge in chiral active systems~\cite{caprini2026modeling}, when two rotating objects interact in a fluid and are subject to hydrodynamic interactions, or when two granular particles experience rotational friction during a collision.
In a homogeneous, isotropic state, the central forces contribute to the pressure, while the transverse forces generate the torque density. The resulting homogeneous virial stress $\bm\Sigma^{({\rm h})}$ has the form~\cite{maire2026Microscopic}
\begin{equation}
\bm\Sigma^{({\rm h})}=-\underbrace{\left(nT+\frac{\pi n^2}{2}\int_0^\infty dr r^2f_{\parallel}(r)g(r;\rho)\right)}_{p(\rho)}\bm 1
+\underbrace{\left(-\frac{\pi n^2}{2}\int_0^\infty dr r^2f_{\perp}(r)g(r;\rho)\right)}_{\tau(\rho)}\bm\varepsilon\,,
\label{eq:tau_virial}
\end{equation}
where $n=\rho/m_p$ is the particle number density. The pair correlation function $g(r;\rho)$ is evaluated at the local density $\rho$, averaged over scales larger than the interaction range but smaller than those of hydrodynamic variations. At fixed temperature, the pressure contains a kinetic contribution already at $\mathcal O(\rho)$, whereas the torque density is purely interaction-driven and therefore starts at $\mathcal O(\rho^2)$. Beyond these leading orders, both $p(\rho)$ and $\tau(\rho)$ acquire a nontrivial density dependence through the pair correlation function $g(r;\rho)$ (see Appendix~\ref{app:transport}).

The simultaneous presence of torque density and odd viscosity in the hydrodynamic equations~\eqref{eq:odd-navier_stokes} can generate a linear instability of the homogeneous state in the compressible regime~\cite{Marconi2026hydrodynamics}, which can explain an immediate formation of bubbles. Additionally, an approximate solution of the nonlinear hydrodynamic equations, obtained by balancing the torque density and pressure contributions, qualitatively reproduces the phase diagram observed in simulations~\cite{Marconi2026hydrodynamics}.
However, these theoretical approaches cannot capture the regime in which bubbles form through nucleation, as occurs in our system.

\subsection{Rayleigh--Plesset equation for a chirality-induced bubble}
\label{sec:void_rayleigh_plesset}

We now ask a simple question: if a circular bubble, namely a void region in the fluid, is created, does it close or grow? The incompressible exterior fluid occupies $r>R(t)$ with uniform density $\rho$ and hence constant torque density $\tau(\rho)=\tau$, while $r<R(t)$ is empty. We assume a sharp circular interface of line tension $\Upsilon$, axisymmetry, and spatially uniform transport coefficients in the surrounding fluid\footnote{An odd tension exerting a tangential force at the interface could also be included, representing the interfacial effect of chiral Korteweg-type stresses~\cite{langford2024phase}.}. 
In a passive fluid at finite temperature, pressure tends to close the void. In a chiral fluid, the torque density $\tau$ generates a transverse force at the interface, driving edge currents whose centrifugal forces oppose pressure and can expand the bubble.

We seek an axisymmetric solution of the odd Navier--Stokes equation outside the circular cavity, expressed in terms of the radial and azimuthal components of the velocity field, $u_r(r,t)$ and $u_\theta(r,t)$, respectively, where $r$ and $\theta$ are the radial and azimuthal coordinates measured from the center of the bubble. The azimuthal flow describes the circulating edge current, while the radial flow accompanies the expansion or collapse of the cavity. 
Incompressibility, axisymmetry, and the kinematic boundary condition $u_r(r=R(t),t)=dR/dt$ allow us to express the radial component of the velocity field as (see Appendix~\ref{app:fluctuating_rayleigh_plesset})
\begin{equation}
u_r(r,t)=\frac{R(t)\dot R(t)}{r}, \qquad r>R(t),
\label{eq:void_radial_velocity}
\end{equation}
where $R(t)$ is the radius of the cavity. Solving for the azimuthal flow and imposing the relevant boundary condition yields, in the slow, overdamped limit, an effective equation for $R(t)$ (see Appendix~\ref{app:fluctuating_rayleigh_plesset} for details on the derivation):
\begin{equation}
\mathcal Z(R)\dot R=F[R]\,,
\label{eq:coupled_rayleigh_plesset}
\end{equation}
where $\mathcal Z(R)>0$ is an effective positive friction coefficient setting the rate of evolution, while $F(R)$ corresponds to a net radial force, which determines whether the bubble expands or collapses, with the following expression
\begin{equation}
F(R)=2\pi R\left(\frac{\rho V_R^2}{2}\left[1-\left(\frac{K_0(R/\ell_\gamma)}{K_1(R/\ell_\gamma)}\right)^2\right]-p_\infty-\frac{\Upsilon}{R}-\frac{2\eta_oV_R}{R}\right)\,.
\label{eq:stationary_void_force}
\end{equation}
Here, $K_n$ is a modified Bessel function of the second kind, of order $n$, which decays exponentially on large scales, and $V_R=u_\theta(R^+)$ is the azimuthal velocity at the cavity boundary:
\begin{equation}
u_\theta(r)=V_R\frac{K_1(r/\ell_\gamma)}{K_1(R/\ell_\gamma)}\,,  \qquad V_R=-\dfrac{\tau\ell_\gamma}{\eta}\dfrac{K_1(R/\ell_\gamma)}{K_2(R/\ell_\gamma)}\,,  \qquad \ell_\gamma=\sqrt{\frac{\eta}{\gamma\rho}}\,.
\label{eq:stationary_void_vortex}
\end{equation}
In Eq.~\eqref{eq:stationary_void_force}, the pressure $p_\infty$ and line tension $\Upsilon$ favor the bubble collapse, while the centrifugal force generated by the edge current favors the bubble expansion, in agreement with the scaling derived in Ref.~\onlinecite{caprini2025Bubble}. Moreover, odd viscosity modifies the normal stress at the interface, as expected from the linear stability analysis developed in Ref.~\onlinecite{Marconi2026hydrodynamics}: its contribution favors expansion when the odd viscosity and torque density are aligned, $\eta_o\tau>0$, while it promotes collapse when $\eta_o\tau<0$.
In what follows, we set $\Upsilon=0$ to isolate the effects of chirality. A positive line tension adds a closing contribution $-\Upsilon/R$ to the radial force per unit circumference, suppressing the growth of sufficiently small cavities. 
We note that when odd viscosity $\eta_o$ and line tension $\Upsilon$ are set to zero, the third and fourth terms on the right-hand side of Eq.~\eqref{eq:stationary_void_force} vanish, and we can recover Eq.~\eqref{eq:simplfiied} with the function $g(x)$ given by:
\begin{equation}
    g(x)=\frac{K_1(x)^2-K_0(x)^2}{K_2(x)^2}\,,
    \label{eq:g}
\end{equation}
with $x=R/\ell_\gamma$.

\subsection{Mechanical work out of equilibrium}

After introducing the net radial force $F(R)$, we define the mechanical work of forming a bubble by integrating the opposing force as the cavity radius increases from zero to $R$:
\begin{equation}
W(R)=-\int_0^R F(R') dR'\,, 
\label{eq:void_mechanical_work}
\end{equation}
with $W(0)=0$, or, equivalently, $F(R)=-dW/dR$.
The function $W(R)$ can be interpreted as an effective potential governing the radial dynamics of the bubble. Its stationary points, determined by the condition $F(R)=0$, correspond to possible steady states, whose stability is determined by the local curvature of $W(R)$. Local minima correspond to dynamically (meta)stable states, whereas local maxima correspond to unstable stationary states and act as barriers separating distinct stable states.
In the present case, $W(R)$ has at most one local maximum, located at $R=R_c$, and at most two local minima. The first minimum, at $R=0$, corresponds to the homogeneous phase, while the second minimum, at $R=R_b$, corresponds to a finite-sized circular bubble of radius $R_b$. This effective-potential picture therefore provides a complete characterization of the reduced, quasistatic radial dynamics.

Transitions from the homogeneous configuration to the bubble state require fluctuations large enough to overcome the barrier at $R=R_c$. We therefore refer to the homogeneous configuration as metastable, in the sense that bubble formation is possible but requires a finite fluctuation. This terminology does not imply which configuration is more frequently observed at long times or in an infinite system.
At equilibrium, the (quasi)probability of observing a critical configuration is controlled by the Gibbs weight, $P(R_c)=\exp[-W(R_c)/T]/Z$, with $Z$ ensuring normalization. Out of equilibrium, however, this expression for the probability is generally no longer valid. Whether such a picture---entirely relying on the critical point of the work function $W$ and the kinetic temperature---remains valid is therefore not obvious, since strong circulating edge currents and other non-equilibrium contributions can potentially facilitate or hinder the nucleation process relative to the equilibrium expectation. Nevertheless, one may still expect the kinetic temperature $T$ to set the typical scale of fluctuations, suggesting an equilibrium-like activation picture in which reaching the critical radius requires fluctuations capable of overcoming a mechanical barrier of order $W(R_c)$.

To assess the validity of this equilibrium-like form, we therefore introduce fluctuations into our odd Navier--Stokes equations, Eq.~\eqref{eq:odd-navier_stokes}, and derive the resulting stochastic dynamics for $R(t)$. This reduction requires several approximations and assumptions, detailed in Appendix~\ref{app:fluctuating_rayleigh_plesset}, which allow us to derive an approximate expression for the probability of finding a cavity of radius $R$ in our non-equilibrium system:
\begin{equation}
P(R)\asymp\exp\left[-\Phi(R)/T\right]\,,  \qquad \Phi(R_c)=\mathcal O \left(W(R_c)\right)\,.
\label{eq:void_quasipotential}
\end{equation}
We find that the probability of observing a given configuration is not directly controlled by the work of formation $W(R)$, but rather by a distinct function $\Phi(R)$. Nevertheless, at the critical radius, $\Phi(R_c)$ and $W(R_c)$ remain of the same order of magnitude. Consequently, $W(R_c)/T$ provides the correct \textit{scale} of the activation exponent, although not its exact value. This observation justifies our study of the work function $W(R)$ even when the system is far from equilibrium.

\section{Theoretical analysis of bubble stability}
\label{sec:3}

\begin{figure}
    \centering
    \includegraphics[width=0.99\linewidth]{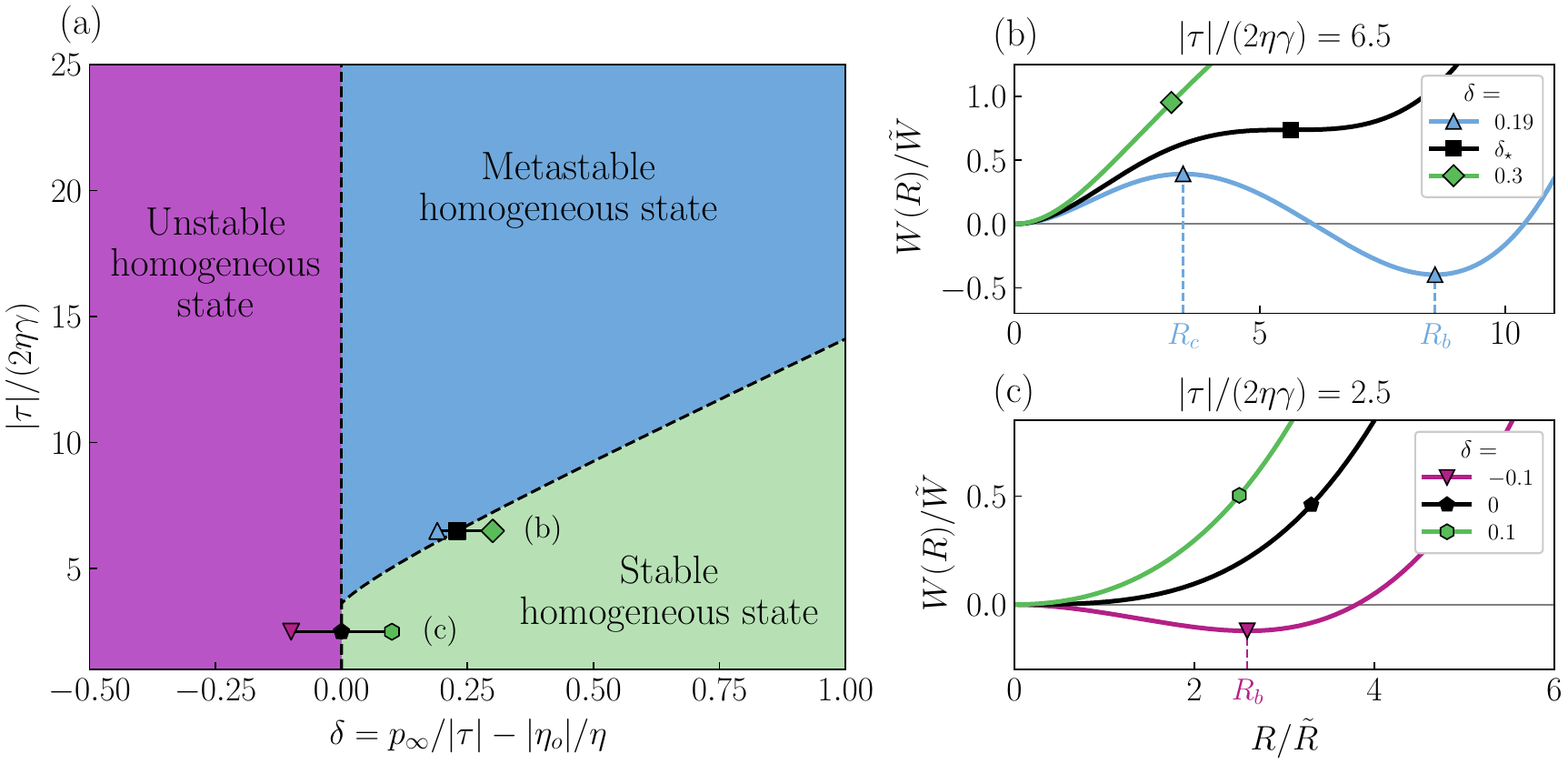}
    \caption{Phase diagram and mechanical work of bubble formation, with odd viscosity and damping at zero line tension ($\Upsilon=0$).  (a) Phase diagram in the plane $(\delta, |\tau|/(2\eta\gamma))$, where $\delta\equiv p_\infty/|\tau|-|\eta_o|/\eta$, at fixed $|\eta_o|/\eta=1$ and $\tau\eta_o>0$. The regions indicate whether the homogeneous state is stable, metastable, or unstable. Each symbol on the horizontal cuts also appears on the corresponding curve in (b) or (c), which shows the mechanical work $W(R)$ obtained from Eqs.~\eqref{eq:stationary_void_force},~\eqref{eq:stationary_void_vortex}, and~\eqref{eq:void_mechanical_work}, with
    $|\tau|/(2\eta\gamma)=6.5$ and $\delta=\{0.19,\delta_\star,0.3\}$ in (b). At the boundary $\delta_\star\simeq0.229923$, the finite-radius maximum and minimum merge. In (c), $|\tau|/(2\eta\gamma)=2.5$ and $\delta=\{-0.1,0,0.1\}$. We defined $\tilde R=\eta/\sqrt{\rho|\tau|}$ and $\tilde W=2\pi|\tau|\tilde R^2$.}
    \label{fig:full}
\end{figure}

A circular bubble of radius $R$ evolves through Eq.~\eqref{eq:coupled_rayleigh_plesset} with the expression for $F(R)$ given by Eq.~\eqref{eq:stationary_void_force}.
For $R>0$, the immediate change in the size of a bubble is equivalently determined by the sign of $F$ or by the shape of $W$, which can be used as a criterion to characterize the stability of the homogeneous state.

In qualitative agreement with the simulation results shown in Fig.~\ref{fig:overview}(c), the theory predicts regions of stability, metastability, and instability of the homogeneous state. These three different states are summarized in Fig.~\ref{fig:full}(a) as a function of two dimensionless parameters, $|\tau|/(2\eta\gamma)$, which measures chiral driving relative to viscous and substrate damping, and $\delta=p_\infty/|\tau|-|\eta_o|/\eta$, which compares pressure with torque density and includes the opening effect of the bubble due to odd viscosity when $\tau\eta_o>0$.
Typical effective work $W(R)$ as a function of $R$ is reported in Fig.~\ref{fig:full}(b) and Fig.~\ref{fig:full}(c) to show the different states that we now discuss in detail:
\begin{itemize}
    \item \textbf{Stable homogeneous state --} 
    At weak chiral driving ($|\tau|/2\gamma \eta\ll 1$) and strong pressure or damping ($\delta \gg 1)$, the effective work $W(R)$ increases monotonically from its minimum $W(0)$ at $R=0$ (see green curves in Fig.~\ref{fig:full}(b) and Fig.~\ref{fig:full}(c)). The edge currents generate insufficient centrifugal force to sustain a bubble, irrespective of its size. Hence, the homogeneous state with $R=0$ is the only stable state (green region in Fig.~\ref{fig:full}(a)).
    
    \item \textbf{Metastable homogeneous state --} By increasing the chiral driving $|\tau|/(2\gamma \eta)$ at fixed, positive $\delta$, sufficiently large cavities can grow (blue region in Fig.~\ref{fig:full}(a)). 
    Indeed, in this regime, the work function $W(R)$ has a maximum at $R_c$ and a minimum at $R_b$ smaller than zero (blue curves in Fig.~\ref{fig:full}(b)). 
    Consequently, perfectly circular cavities smaller than $R_c$ collapse to $R=0$, while those larger than $R_c$ approach the stable radius $R_b$.
    In the weak-damping limit at fixed transport coefficients and $\eta p_\infty>\tau\eta_o$, the critical radius $R_c$ can be approximated as 
    \begin{equation}
        R_c\simeq\sqrt{\frac{8\eta(\eta p_\infty-\tau\eta_o)}{\rho\tau^2}}\,,  
        \label{eq:weak_damping_critical_radius}
    \end{equation}
    while the stable $R_b$ has the following expression:
    \begin{equation}
        R_b\simeq\frac{\tau^2}{2p_\infty\sqrt{\rho\eta\gamma^3}}\,.
        \label{eq:weak_damping_radii}
    \end{equation}
    These estimates show that increasing pressure or shear viscosity, with other parameters fixed, raises the critical radius $R_c$ but reduces the stable radius $R_b$. Stronger damping also reduces $R_b$, whereas $R_c$ is independent of damping to leading order. As $\gamma\to0$, $R_b\propto\gamma^{-3/2}$ diverges, so no finite bubble size is selected (see Appendix~\ref{app:stability} for details). Without damping, the edge-current speed grows with the cavity radius, so the centrifugal contribution, proportional to $u_\theta^2$, eventually overcomes pressure for a sufficiently large cavity, even at finite viscosity.
    
    \item \textbf{Unstable homogeneous state --} Odd viscosity can eliminate the nucleation barrier, allowing bubble formation to proceed without activation. For $\delta<0$, equivalently $\eta p_\infty<\tau\eta_o$, arbitrarily small bubbles grow, and the homogeneous state becomes unstable (violet region in Fig.~\ref{fig:full}(a)). In this regime, the critical radius $R_c$ in Eq.~\eqref{eq:weak_damping_critical_radius} ceases to exist, so $W(R)$ has no barrier and decreases monotonically toward the stable radius $R_b$ in Eq.~\eqref{eq:weak_damping_radii} (violet curve in Fig.~\ref{fig:full}(c)).
    This is the nonlinear realization of the linear instability found in Ref.~\onlinecite{Marconi2026hydrodynamics}. Indeed, the condition $p_\infty < \tau\eta_o/\eta $ indicates that the odd-viscous normal stress induced by torque-driven shear overcomes the closing pressure. 
    Odd viscosity thus controls the onset of growth at small $R$, while centrifugal forces and damping still select a finite bubble radius $R_b$.
\end{itemize}
We finally discuss two limiting cases that can be treated analytically: vanishing damping, $\gamma\to0$, and zero odd viscosity, $\eta_o=0$. The corresponding derivations and counterparts of Fig.~\ref{fig:full} are presented in Appendix~\ref{app:stability}, while we only summarize the main results here.

\begin{figure}[t]
    \centering
    \includegraphics[width=0.85\linewidth]{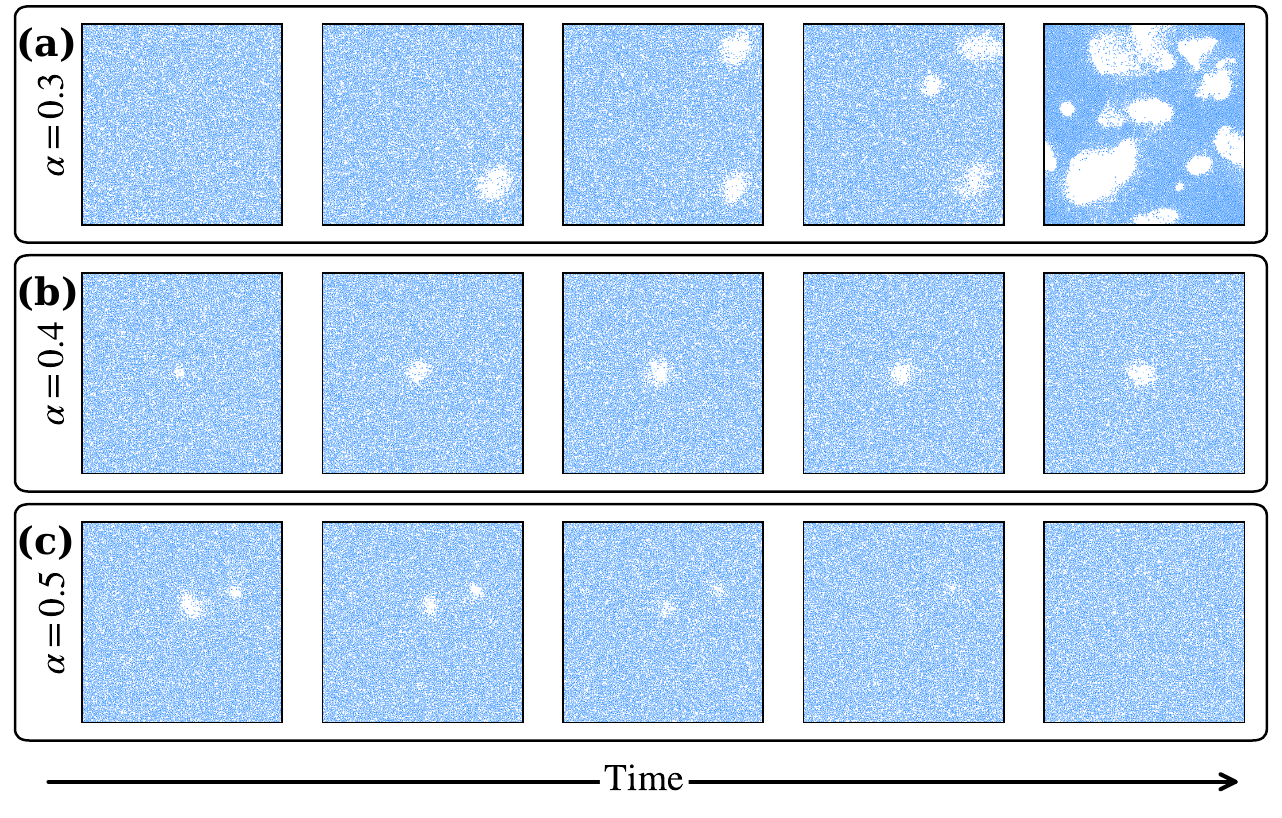}
    \caption{Time evolution at $\phi=0.45$ and $N=2\times 10^4$, with time increasing from left to right. (a) At $\alpha=0.3$, bubbles form without an initial seed, and the system evolves towards a chaotic state; $t\Delta/\sigma=\{100,1000,1400,1800,4500\}$. (b) At $\alpha=0.4$, a seeded bubble reaches a finite size and persists; $t\Delta/\sigma=\{0,300,1250,2000,10000\}$. (c) At $\alpha=0.5$, a seeded cavity collapses and the homogeneous state is recovered; $t\Delta/\sigma=\{0,100,200,300,400\}$. Throughout $\gamma\sigma/|\Delta|=0.05$.}
    \label{fig:time_evol}
\end{figure}

\vskip5pt
\noindent
\textbf{Theory for vanishing odd viscosity --} With $\eta_o=0$, the homogeneous state can never be unstable and only regions of stability or metastability can occur. Since in this case $\delta>0$, the phase diagram is consistent with the right part of Fig.~\ref{fig:full}(a) (see Fig.~\ref{fig:noeta} in Appendix~\ref{app:stability}).
Anticipating the comparison with simulations, we note that a true instability is difficult to distinguish numerically from weak metastability, namely a regime with a small nucleation barrier, since both lead to rapid bubble formation. Moreover, the absence of an independent estimate of $\eta_o$ prevents a quantitative placement of the simulations within the phase diagram. Consequently, in the overview figure for our results (Fig.~\ref{fig:overview}), we have reported the theoretical prediction for $\eta_o=0$ for simplicity.

\vskip5pt
\noindent
\textbf{Theory for vanishing damping --}
The vanishing-damping case, $\gamma=0$, constitutes a singular limit in which the homogeneous state is never stable and no finite-size bubble exists since $W(R\to \infty)\to -\infty$. Physically, this follows from the different scaling of the competing forces with the bubble radius. While the closing pressure traction does not grow with $R$, the centrifugal traction does, so that there always exists a sufficiently large radius beyond which centrifugal effects dominate and drive further expansion. Accordingly, a phase diagram, in the plane of odd viscosity $\eta_o$ and torque density $\tau$, shows only metastable and unstable regions (see Fig.~\ref{fig:nodamping} in Appendix~\ref{app:stability}).

\section{Further numerical evidence of metastability in chiral fluids and qualitative agreement with theory}
\label{Sec:Numerics}

\begin{figure}[t]
    \centering
    \includegraphics[width=0.99\linewidth]{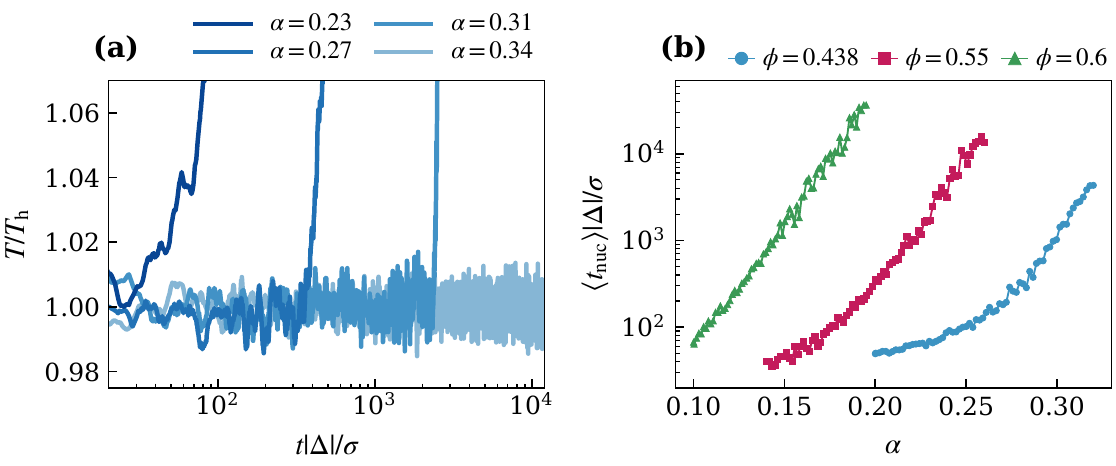}
    \caption{(a) 
    Time evolution of the global kinetic temperature $T = m_p\langle \bm{v}^2\rangle/2$ for different values of the restitution coefficient $\alpha=0.23$, $0.27$, $0.31$, and $0.34$---the smaller $\alpha$, the larger the chirality---at fixed packing fraction $\phi=0.438$. We normalize the temperature $T$ by the kinetic temperature in the corresponding homogeneous phase, $T_{\text{h}}$, so $T/T_{\text{h}}$ fluctuates around 1 before nucleation events occur. (b) Mean nucleation time $\langle t_{\text{nuc}}\rangle$, normalized by the time $\sigma/\Delta$, as a function of $\alpha$ for $\phi=0.6$, $0.55$, and $0.438$. The higher $\alpha$ (lower chirality), the larger the nucleation time.
    For the system sizes considered, $t_{\rm nuc}$ is defined as the first time the system temperature exceeds $1.07$ times the expected homogeneous temperature.
    In (b), each point represents an average over 20–150 independent simulations.
    The other simulation parameters are $N=10^4$ and $\gamma\sigma/|\Delta|=0.05$.
    } 
    \label{fig:dynamics}
\end{figure}
Having derived a theory establishing the conditions for stability, metastability, and instability of the homogeneous state of a chiral active fluid, we now discuss the simulation results partially anticipated in Sec.~\ref{sec:overview}. In this section, we investigate in more detail the time evolution of bubbles and provide further evidence for metastability by monitoring the kinetic temperature and measuring the nucleation time.

As anticipated in Fig.~\ref{fig:overview}(c), the homogeneous state can be stable, metastable, or unstable, depending on the packing fraction value $\phi$ and chirality, through the restitution coefficient $\alpha$: the larger the value of $\alpha$, the lower the chirality.
Even if, in simulations, it is not easy to distinguish between metastable and unstable homogeneous states, we note that at relatively small $\alpha$ (large chirality) and high $\phi$, bubbles develop almost immediately, even by starting from a homogeneous state (grey region in Fig.~\ref{fig:overview}(c)). A typical time evolution in this regime is reported in Fig.~\ref{fig:time_evol}(a).
As alpha increases, the waiting time grows until no bubbles are observed within the simulation time (pink region in Fig.~\ref{fig:overview}(c)). However, in this regime, we observe bubbles growing and reaching a finite steady-state size if we start the simulation from an inhomogeneous initial configuration, namely a circular void, termed a seed (see the time evolution in Fig.~\ref{fig:time_evol}(b)).

To systematically analyze bubble nucleation, we have monitored the time evolution of the kinetic temperature $T$, starting from a homogeneous state with temperature $T_{\text{h}}$. This analysis has been performed for different values of the restitution coefficient $\alpha$ (Figure~\ref{fig:dynamics}(a)).
The normalized temperature $T/T_{\text{h}}$ is used as a proxy for nucleation since the formation of a bubble coincides with the generation of edge currents, which efficiently inject energy into the system and increase the global kinetic temperature \cite{caprini2025Bubble}.
At the smallest $\alpha$, the kinetic temperature $T$ immediately increases, indicating the almost instantaneous formation of a bubble. For $\alpha=0.27$ and $0.31$, the same sharp increase occurs only after a finite waiting time, which we identify with the nucleation time. For $\alpha=0.34$, the temperature remains close to its initial level throughout the recorded trajectory. 
This nucleation time $t_{\text{nuc}}$ extracted from the kinetic temperature time evolution is studied as a function of $\alpha$ (Fig.~\ref{fig:dynamics}(b)) for different packing fraction values after averaging over several realizations. 
A small increase in $\alpha$, i.e., a decrease in chirality, increases the nucleation time by up to two orders of magnitude, until spontaneous nucleation is no longer observed within the simulation time and seeding is required to still observe bubbles, as in Fig.~\ref{fig:overview}(c). This sharp increase is consistent with our theoretical description in terms of nucleation. Decreasing chirality raises the nucleation barrier and therefore \emph{exponentially} suppresses the probability of crossing it, provided that the kinetic temperature does not vary significantly.

\section{Finite size and shape stability of an isolated bubble}
\label{sec:4}

\begin{figure}
    \centering
    \includegraphics[width=0.99\linewidth]{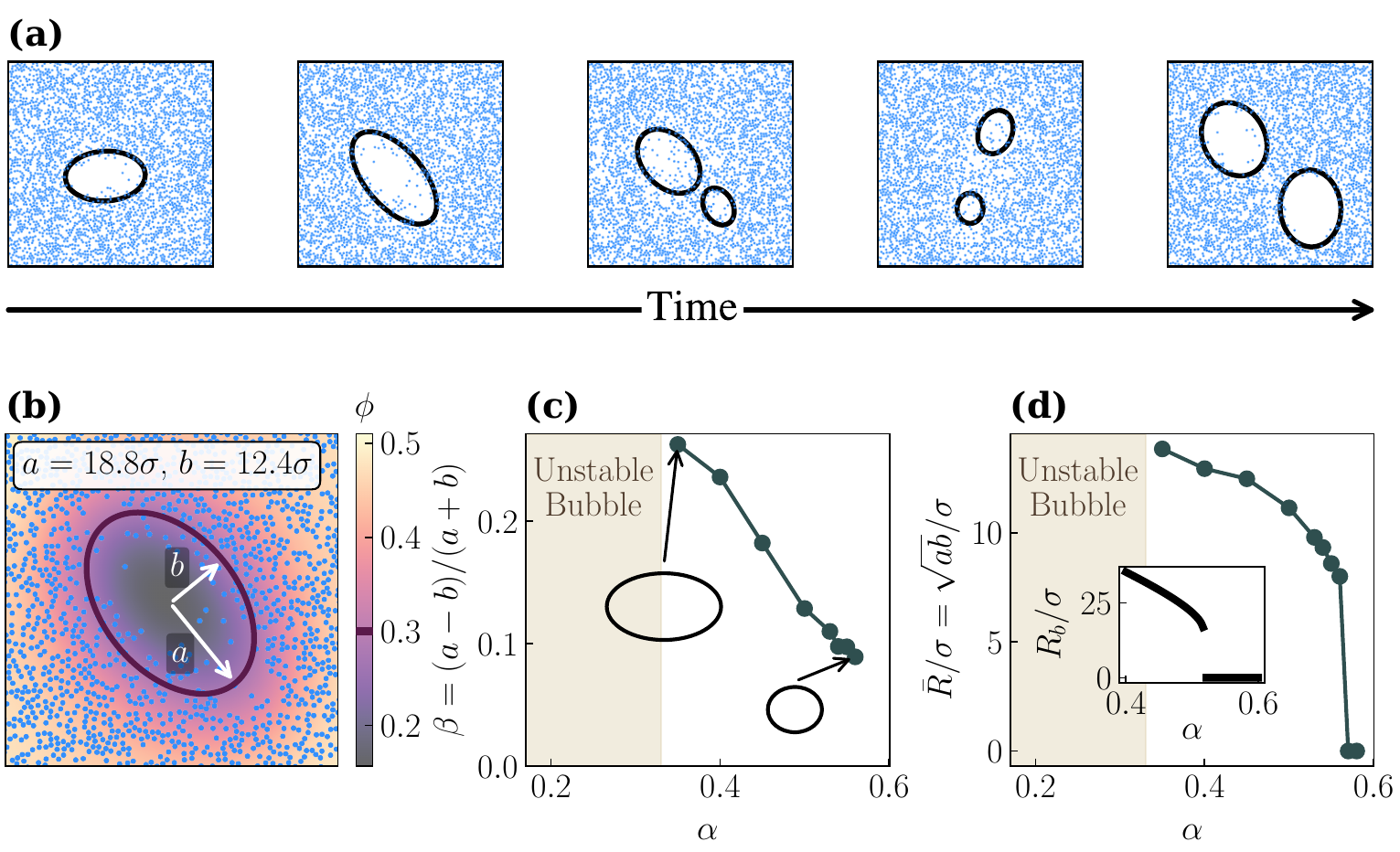}
    \caption{Bubble deformation, size, and asymmetry from seeded configurations at $\phi=0.5$ and $N=3\times 10^4$. (a) Elongation and splitting of an initially approximately circular bubble at $\alpha=0.3$ (zoomed views). From left to right, $t\Delta/\sigma=\{150,200,250,300,350\}$; black outlines show ellipse fits to the bubbles. (b) We smooth the density field and extract a closed isodensity contour at $\phi^{\rm threshold}_{\rm smooth}=0.3$, then fit an ellipse with semi-axes $a$ and $b$. (c) Averaged asymmetry of a single bubble $\beta=(a-b)/(a+b)$ as a function of $\alpha$. (d) Averaged effective radius $\bar R=\sqrt{ab}$ as a function of $\alpha$. Averages in (c,d) are performed over 100 snapshots for each point. Throughout $\gamma\sigma/|\Delta|=0.05$.}
    \label{fig:bubblesize}
\end{figure}

A remaining prediction of our theory is the existence of a finite bubble size $R_b$ after nucleation. Indeed, as shown in Fig.~\ref{fig:time_evol}(b) for $\alpha = 0.4$, a seeded configuration can grow to a finite size and remain as an isolated bubble throughout the observation time. Typically, in this regime, we need to wait a long simulation time to observe a bubble starting from a homogeneous state. At larger chirality, $\alpha=0.3$ in Fig.~\ref{fig:time_evol}(a), bubbles form spontaneously from the homogeneous state, without an imposed seed. However, in this regime of stronger chirality, no stable bubble is found, and the system eventually reaches a chaotic state in which many bubbles repeatedly form, deform, merge, and break apart. This long-time state has been studied in Ref.~\onlinecite{petrini2026sparkling} and requires a many-body description beyond our analysis. Here, we focus instead on the early-time dynamics.

To better understand these findings, we examine the size and shape of an isolated bubble as the chirality is varied through $\alpha$. Fig.~\ref{fig:bubblesize}(a) shows the early-time dynamics of a state that ultimately becomes chaotic. Although the initially formed bubble first approaches a well-defined finite size, it subsequently elongates and splits into two bubbles. These daughter bubbles eventually undergo further splitting, leading to the many-bubble state shown in Fig.~\ref{fig:time_evol}(a). Beyond this splitting, additional nucleation far from the already formed bubbles can also arise. From these results, it appears that the shape of the bubble plays an important role. To analyze it, we extract a closed isodensity contour from the smoothed density field and fit an ellipse with major and minor semi-axes $a$ and $b$, as illustrated in Fig.~\ref{fig:bubblesize}(b). We define the effective radius $\bar R=\sqrt{ab}$ as the radius of a circle with the same area, and the asymmetry $\beta=(a-b)/(a+b)$ as a measure of elongation. Figs.~\ref{fig:bubblesize}(c,d) show that, within the regime of persistent bubbles, decreasing $\alpha$ increases both size and elongation, eventually leading to breakup. Increasing $\alpha$ instead produces smaller, more circular bubbles, until the seed collapses. At this boundary, the persistent bubble disappears with a finite jump in radius, consistent with the theoretical prediction at $\delta_*$ in Fig.~\ref{fig:full}(b). The inset of Fig.~\ref{fig:bubblesize}(d) illustrates this qualitative agreement, although the predicted transition location is not quantitatively accurate. Thus, isolated bubbles persist over an intermediate range of $\alpha$, becoming non-circular toward its lower end before breaking apart, and collapsing at its upper end.

In Appendix~\ref{app:noncircular}, we perturb the circular boundary at $R_b$ by a small displacement $\delta r(\theta,t)=\mathrm{Re}[h e^{st+im\theta}]$ and calculate its coupled evolution with the surrounding flow. Here, $m\ge2$ labels the shape deformation, with $m=2$ corresponding to elongation, and $\mathrm{Re}[s]>0$ means that its amplitude grows. We find that, theoretically, for all state points that can form a bubble with a finite size $R_b$, this bubble is unstable to at least one non-circular $m\geq 2$ mode. The largest growth rate decreases toward the boundary of bubble metastability but remains positive. Slower growth near this boundary is compatible with deformations that saturate into a persistent non-circular bubble, as observed in Fig.~\ref{fig:bubblesize}(c). Faster growth deeper in the nucleation region may instead favor the elongation and splitting seen in Fig.~\ref{fig:bubblesize}(a). This interpretation remains tentative, since the linear calculation predicts the initial growth of deformations, but not their final shape or the threshold for breakup.

\section{Discussion}
\label{sec:5}

Here, we have developed a nucleation theory for chiral active fluids, which inject angular momentum during collisions and display the spontaneous formation of bubbles, even in the absence of attractions.
Our theory and simulations predict the existence of regimes depending on the chirality value and density where the homogeneous state is stable (no bubbles), metastable (bubbles need a typical time to nucleate), unstable (bubbles form immediately). In analogy with the equilibrium picture, we are able to derive effective dynamics for the bubble radius and to predict the stable and critical radius, as a function of the model parameters. 
In addition, we investigated the shape stability of an isolated bubble to understand the instability of circular bubbles observed in our system.

The first unresolved question is the nonlinear stability of an isolated bubble. The growing shape modes found in Appendix~\ref{app:noncircular} show that the circular solution cannot describe the late dynamics throughout the sampled nucleation region. Their growth rates alone do not establish whether the bubble settles into a stable ellipse or breaks apart. Resolving this requires following the deformed interface beyond the linear regime. Related studies provide examples of persistent elongated shapes and division in active nematic droplets~\cite{giomi2014division}, chemically driven droplet division~\cite{zwicker2017division}, and breakup of elongated domains in chiral systems~\cite{burekovic2026bulk}.
Whether this instability underlies the sparkling phase reported in Ref.~\onlinecite{petrini2026sparkling} is unclear. In this chaotic state, also seen in Fig.~\ref{fig:time_evol}(a), bubbles continually form, deform, merge, and break apart. The rapid growth of isolated-bubble deformations suggests a possible connection, but sparkling may be primarily collective. Flows generated by neighboring bubbles could strain their interfaces too strongly for persistent shapes to survive. Local densification between bubbles could further amplify these flows through the density dependence of the torque.

Stable arrays of bubbles have been reported in other chiral systems~\cite{digregorio2025phase,caprini2025Bubble}. The exterior circulating flow around a circular void in an incompressible chiral fluid suggests an analogy with a spinning disk in a passive fluid. In the latter system, long-ranged Magnus repulsion stabilizes hexatic order, which gives way to chaotic motion at higher Reynolds number~\cite{goto2015hydrodynamic}. Similar interactions likely organize bubbles into a lattice in the system with chiral particles. Conversely, loss of this order at higher chirality or Reynolds number could promote close encounters, deformation, and merging, providing a collective route to sparkling that need not originate in the shape instability of an isolated bubble.
The approach to either collective state also requires a theory of bubble growth and coarsening. A description of $N_b$ bubbles, coupled through their positions, sizes, and the surrounding density field, could follow the self-consistent treatment of supersaturation in Lifshitz--Slyozov--Wagner theory~\cite{bray1994theory}. Such a theory would need to identify what controls material redistribution and growth competition in place of an equilibrium chemical potential. It would also clarify whether the characteristic bubble size is set by damping through the isolated-bubble radius $R_b$, or by collective mechanisms that arrest coarsening, such as those associated with effective negative interfacial tensions in active phase separation~\cite{cates2024active}. 

\begin{figure}[!t]
    \centering
    \includegraphics[width=0.99\linewidth]{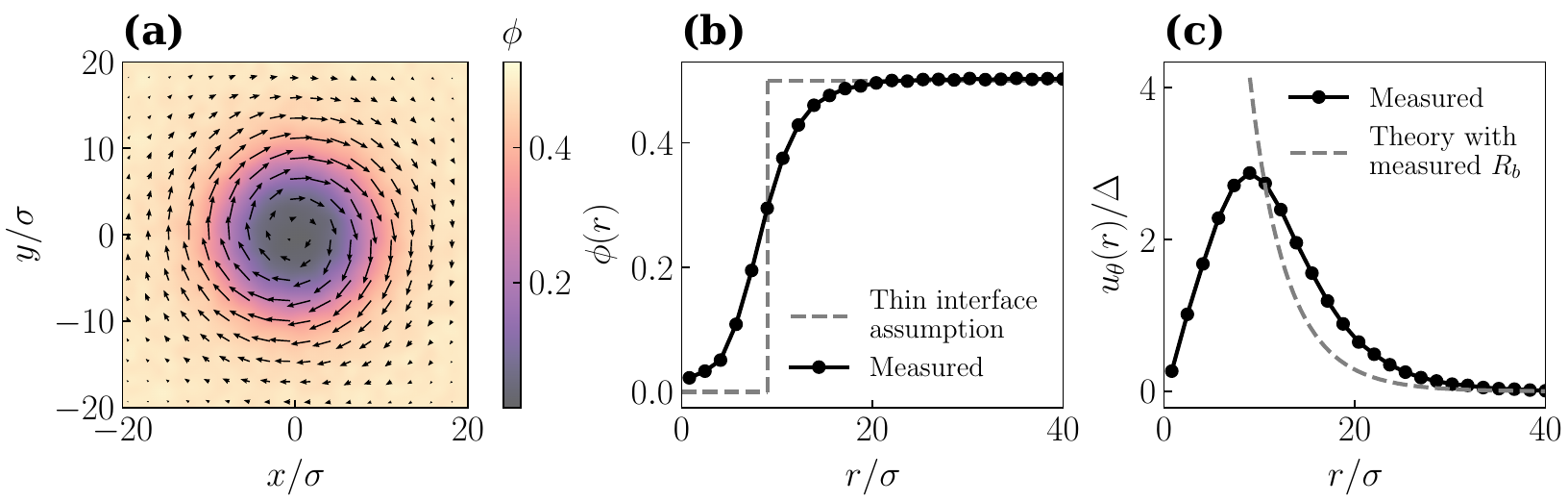}
    \caption{Measured density and velocity around an isolated bubble at $N=3\times 10^4$, $\alpha=0.56$, and $\phi=0.5$. (a) Density map and velocity field averaged over 761 snapshots after centering on the vortex. (b) Radially averaged density profile. The dashed step illustrates the sharp-interface approximation, with its position set by the measured velocity maximum. (c) Measured azimuthal velocity profile and the theoretical exterior flow from Eq.~\eqref{eq:stationary_void_vortex}, using the measured radius $R_b=9\sigma$ (dashed, $r\geq R_b$).}
    \label{fig:profiles}
\end{figure}

A more quantitative description must also resolve the interface and its surrounding fields. Figure~\ref{fig:profiles} shows an isolated bubble at $\alpha=0.56$ and $\phi=0.5$, averaged over many configurations centered on the cavity. The edge current is concentrated near the cavity boundary, but both density and azimuthal velocity vary smoothly across it. The density transition has a width $w$ that can be comparable to the stable or critical bubble radius, with residual ellipticity further broadening the radial average. Density-gradient stresses allow such smooth interfaces~\cite{anderson1998diffuse}, represented approximately by $\phi(r)\simeq\phi(r\to\infty)(1+\tanh[(r-R)/w])/2$ with our sharp density step recovered as $w\to0$. The exterior-flow prediction in Fig.~\ref{fig:profiles}(c), evaluated at $R_b=9\sigma$, therefore provides only an approximate comparison across this diffuse boundary.

These profiles motivate a compressible hydrodynamic description~\cite{zhang2023unified} in terms of coupled density, velocity, and temperature fields. A natural extension of the present framework would combine the continuity and momentum equations with an equation of state $p=p(\rho,T)$ and capillary Korteweg stresses~\cite{maire2026Microscopic}, following the compressible approach developed in Ref.~\onlinecite{Marconi2026hydrodynamics}. The latter could be further extended to include a temperature equation accounting for energy injection, transport, and dissipation. Such a formulation would distribute the azimuthal driving across the interface while retaining the viscous contributions associated with velocity-field divergence, thereby modifying both the local force balance and the hydrodynamic coupling between bubbles.

The present theoretical framework provides a guide for tuning bubble formation in experiments involving spinning colloids \cite{massana2021arrested} and chiral active spinners \cite{scholz2018rotating,liu2020oscillating,vega2022diffusive,caprini2025spontaneous}. In these systems, the relevant control parameters can be varied experimentally, offering a means to steer both the onset and subsequent evolution of the bubble phase. Thus, the effective description developed here establishes a direct connection between experimentally accessible parameters and the conditions for bubble formation, and may be exploited to design protocols for controlling bubble nucleation and growth in chiral active systems.

\section*{Acknowledgments}
LC acknowledges financial support from the University of Rome Sapienza, under the project Ateneo 2024 (RM125199C3E64E09) ‘‘Non-equilibrium behavior and collective response across scales in living and engineered active matter''.

\appendix

\makeatletter
\renewcommand{\@seccntformat}[1]{%
  \ifstrequal{#1}{section}{Appendix~}{}%
  \csname the#1\endcsname\quad
}
\makeatother

\section{Mapping particle simulations to hydrodynamics}
\label{app:transport}

The comparison with particle simulations in Secs.~\ref{sec:2}, \ref{Sec:Numerics} and \ref{sec:4} requires to know $p_\infty$, $\tau$, $\eta$, and $\eta_o$ at each microscopic state point $(\phi,\alpha,\Delta,\gamma)$. We use homogeneous bulk estimates, with number density $n=4\phi/(\pi\sigma^2)$ and mass density $\rho=m_p n$. The coefficients are evaluated at the global density, i.e., we neglect the densification of the surrounding fluid in a finite system as the cavity grows (which holds for $L\gg R_c$ if we are interested in nucleation and $L\gg R_b$ if we are interested in finite size selection). Length, time, and kinetic energy are measured in units of $\sigma$, $\sigma/|\Delta|$, and $m_p\Delta^2$.

\subsection{Homogeneous temperature, pressure, and torque}

\begin{figure}[!ht]
    \centering
    \includegraphics[width=0.9\linewidth]{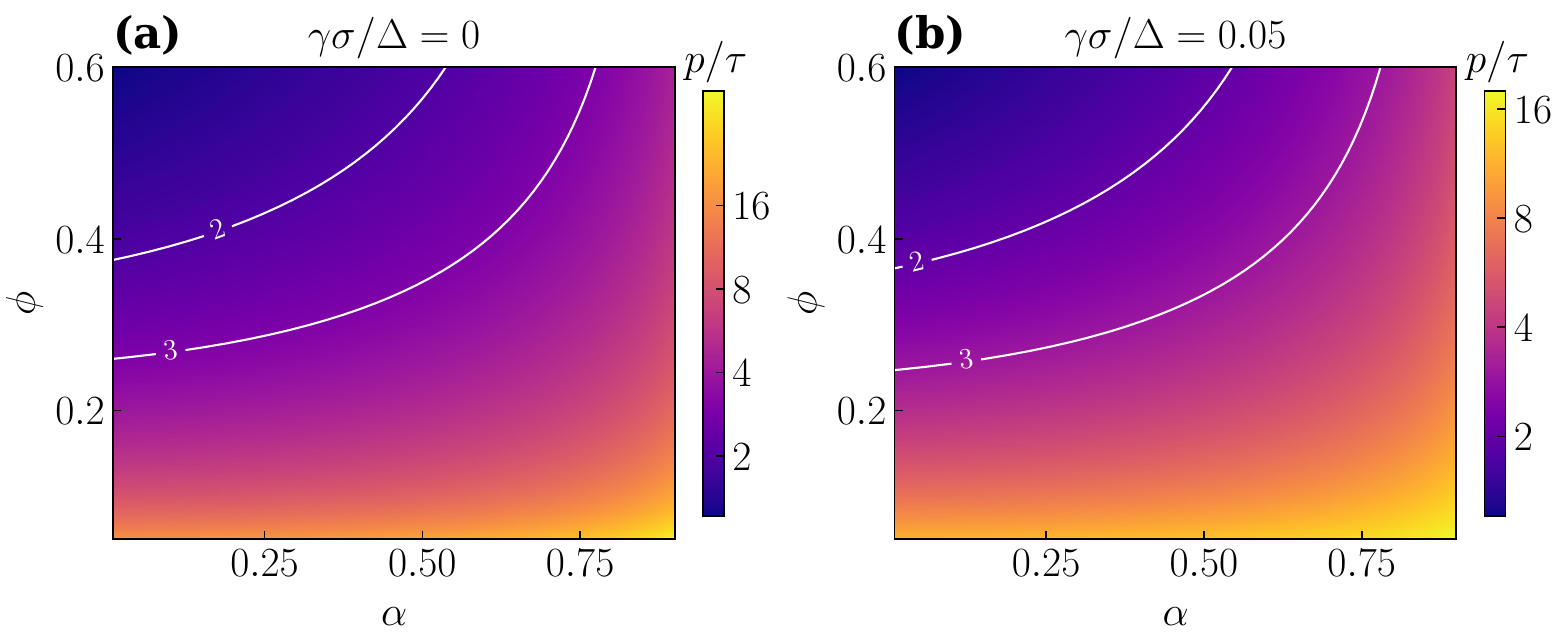}
    \caption{Pressure-to-torque ratio $p_\infty/\tau$ from the homogeneous stress closure, Eq.~\eqref{eq:theoretical_stress}, for (a) $\gamma=0$ and (b) $\gamma\sigma/\Delta=0.05$.}
    \label{fig:torque/pressure}
\end{figure}

The homogeneous temperature balance in the system, assuming a Gaussian velocity distribution, is~\cite{maire2026Microscopic}
\begin{equation}
    \partial_t T=\dfrac{\omega_{\rm coll}(\phi, T)}{2}\langle\delta E \rangle_{\rm coll}-2\gamma T=\sqrt{\pi}\chi\sigma n\sqrt{\frac{T}{m_p}}\Big[m_p\Delta^{2}-(1-\alpha^{2})T\Big]-2\gamma T\,,
    \label{eq:homogeneous_energy_balance}
\end{equation}
where $\langle\delta E\rangle_{\rm coll}=m_p\Delta^2-(1-\alpha^2)T$ is the mean total pair-energy change and $\omega_{\rm coll}=2\sqrt{\pi}\chi\sigma n\sqrt{T/m_p}$ is the collision frequency per particle. $\chi$ is the value of the pair correlation function at $r=\sigma$; we assume it takes its equilibrium hard-disk value: $\chi\simeq(1-7\phi/16-\phi^3/20)/(1-\phi)^2$~\cite{mulero2009equation}. The nonzero stationary solution to Eq.~\eqref{eq:homogeneous_energy_balance} is
\begin{equation}
    T=\frac{m_p\left(-\gamma + \sqrt{\gamma^2+\pi \chi^2\sigma^2 n^2 \Delta^2(1-\alpha^2)}\right)^2}{\pi \chi^2\sigma^2 n^2 (1-\alpha^2)^2}\,.
    \label{eq:steady_state_temperature}
\end{equation}
The bulk stresses are~\cite{maire2026Microscopic}
    \begin{subequations}
    \label{eq:theoretical_stress}
    \begin{flalign}
    p&=n T + n\phi \chi(1+\alpha)T\,,\label{eq:pressure_kinetic}\\
    \tau &= 2n\phi\chi\sqrt{\frac{m_p T}{\pi}}\Delta \,.\label{eq:torque_kinetic}
    \end{flalign}
\end{subequations}
We note a factor-of-$\pi$ typo in Ref.~\onlinecite{maire2026Microscopic} for $\tau$.
We recall that $p_\infty$ is the pressure at infinity, which coincides with the value given by the equation of state Eq.~\eqref{eq:pressure_kinetic} for a given density and temperature. Fig.~\ref{fig:torque/pressure} illustrates the pressure--torque competition: torque becomes comparable to pressure at high density and low $\alpha$, favoring nucleation.

\subsection{Viscosities and limits of the mapping}

Viscosities describe response to gradients and are harder to estimate than homogeneous stresses. Measurements are difficult~\cite{maire2026Microscopic}, especially if the homogeneous state decays before sampling. A reliable closure at $\phi\gtrsim0.1$ is unavailable for this model. We therefore use the Enskog viscosity of granular disks ($\Delta=\gamma=0$), evaluated at the active temperature~\cite{garzo1999dense}:
\begin{equation}
\eta(\phi,T,\alpha)= \frac{1}{2\sigma}\sqrt{\frac{m_p T}{\pi}}
\left[\frac{\left(2+\phi\chi(1+\alpha)\right)\left(4-\phi\chi(1+\alpha)(1-3\alpha)\right)}{\chi(7-3\alpha)(1+\alpha)}+\frac{4}{\pi}\phi^2\chi(1+\alpha)
\right]\,.
\end{equation}
The assumption $\gamma=0$ is well justified in our system, as the damping is always kept small ($m(\sigma\gamma)^2\ll m\Delta^2$). By contrast, the assumption $\Delta=0$ for the viscosity is an uncontrolled approximation introduced to make further progress.

We do not have a large-density estimate for the odd viscosity. The two viscosities are comparable at $\phi=0.01$, but already at $\phi=0.1$ the reported maximum is $\eta_o/\eta\simeq 0.2$. These trends motivate
\begin{equation}
\eta_o=0
\end{equation}
in the particle comparison. Odd viscosity could still change the distinction between metastability and direct instability, as discussed in Sec.~\ref{sec:3}.

\section{Fluctuating Rayleigh--Plesset equation}
\label{app:fluctuating_rayleigh_plesset}

\subsection{Fluctuating chiral Ekman--Navier--Stokes equation}

We start from the odd Navier-Stokes equation (Eq.~\eqref{eq:odd-navier_stokes}) with an additional random collisional stress $\bm\Sigma^{(\rm rand)}$ to account for fluctuations driving nucleation,
\begin{align}
\rho(\partial_t\bm u+\bm u\cdot\bm\nabla\bm u) &=-\gamma\rho\bm u+\bm\nabla\cdot\bm\Sigma^{(\rm tot)}\nonumber\\
&=-\gamma\rho\bm u-\bm\nabla\mathtt p+\eta\bm\nabla^2\bm u +\eta_o\bm\varepsilon\cdot\bm\nabla^2\bm u +\bm\varepsilon\cdot\bm\nabla\tau +\bm\nabla\cdot\bm\Sigma^{(\rm rand)}\,, \qquad \bm\nabla\cdot\bm u=0\,,
\label{eq:appendix_fluctuating_NS}
\end{align}
with the total stress:
\begin{equation}
\bm\Sigma^{(\rm tot)} =-\mathtt p\bm 1+\tau\bm\varepsilon+2\eta\bm D +\eta_o(\bm\varepsilon\cdot\bm D-\bm D\cdot\bm\varepsilon) +\bm\Sigma^{(\rm rand)}\,, \qquad D_{ij}=\dfrac{1}{2}(\partial_i u_j+\partial_j u_i)\,.
\label{eq:added_total_stress}
\end{equation}
Here $\bm\varepsilon$ is the two-dimensional Levi-Civita symbol, and the other constant terms are discussed below Eq.~\eqref{eq:odd-navier_stokes}.
For simplicity, we assume that $\bm\Sigma^{(\rm rand)}$ satisfies a Fluctuation-Dissipation relation:
\begin{equation}
\left\langle\Sigma^{(\rm rand)}_{ij}(\bm r,t) \Sigma^{(\rm rand)}_{kl}(\bm r',t')\right\rangle =2\eta T(\delta_{ik}\delta_{jl}+\delta_{il}\delta_{jk} - \delta_{ij}\delta_{kl})\delta(\bm r-\bm r')\delta(t-t')\,.
\label{eq:appendix_LL_covariance}
\end{equation}
This is usually a good approximation for active granular systems~\cite{maire2025dynamical}. The proper noise can be derived via a fluctuating Boltzmann equation~\cite{bodineau2023statistical, feliachi2022fluctuating}.

\subsection{Kinematic, traction, and far-field boundary conditions}
\label{app:circular_boundary_conditions}

We now want to model a circular cavity of radius $R$. To do so, we follow the standard Rayleigh--Plesset derivation~\cite{van2024soft,van2016dynamics}, adapted to a two-dimensional cavity. 

\paragraph{Polar coordinates and kinematic condition.} We start by expressing the hydrodynamics in polar coordinates $r$ and $\theta$, the radial and tangential coordinates, respectively. After introducing the radial and azimuthal components of the velocity field, $u_r(r,t)$ and $u_\theta(r,t)$, and assuming axisymmetry (no dependence on $\theta$), incompressibility $\bm\nabla \cdot \bm{u}=0$ can be expressed in polar coordinates as 
\begin{equation}
\frac{1}{r}\frac{\partial}{\partial r}\left(r u_r\right)=0 \,.
\label{eq:continuity_app}
\end{equation}
Integrating with respect to $r$ yields $r u_r=C(t)$. By determining $C(t)$ through the boundary condition at $r=R(t)$ and the kinematic condition $u_r(r=R,t)=\dot R$, the radial component of the velocity field reads
\begin{equation}
u_r(r,t)=\frac{R(t)\dot R(t)}{r}\,, \qquad r>R(t)\,,
\label{eq:void_radial_velocity_app}
\end{equation}
which corresponds to Eq.~\eqref{eq:void_radial_velocity}.
From the continuity equation~\eqref{eq:continuity_app} and the kinematic condition $u_r(r=R,t)=\dot R$, we can express the radial derivative of the velocity profile calculated at $r=R$ as
\begin{equation}
    \partial_r u_r(R)=-\frac{\dot{R}}{R}\,.
    \label{eq:expression_radialderivative_velocity}
\end{equation}
This expression will be useful later.

\paragraph{Traction balance at the void.}
Since the void cavity is traction-free, the force balance reduces to
\begin{equation}
\bm\Sigma^{(\rm tot)}(R)\cdot\bm n =\frac{\Upsilon}{R}\bm n\,,
\label{eq:app_vector_traction}
\end{equation}
where $\Upsilon$ corresponds to the line tension and $\bm n=-\bm e_r$ is the normal to the bubble contour. Consequently, the normal and tangential components of the traction are:
\begin{subequations}
    \begin{flalign}
    &\Sigma^{(\rm tot)}_{rr}(R)=\Upsilon/R\\ &\Sigma^{(\rm tot)}_{\theta r}(R)=0. 
\end{flalign}
\label{eq:traction_balance_condition}
\end{subequations}
The second condition follows from the absence of a tangential line force when $\Upsilon$ is constant. The components of the expression~\eqref{eq:added_total_stress} for the total stress $\bm\Sigma^{(\rm tot)}$, for axisymmetric flow, are
\begin{subequations}
\begin{flalign}
\label{eq:added_polar_tractions_1}
\Sigma^{(\rm tot)}_{rr} &=-\mathtt p+2\eta\partial_r u_r +\eta_o\left(\partial_r u_\theta-\frac{u_\theta}{r}\right) +\overline\Sigma^{(\rm rand)}_{rr}\,\\
\label{eq:added_polar_tractions_2}
\Sigma^{(\rm tot)}_{\theta r} &=-\tau+\eta\left(\partial_r u_\theta-\frac{u_\theta}{r}\right) +\eta_o\left(\frac{u_r}{r}-\partial_r u_r\right) +\overline\Sigma^{(\rm rand)}_{\theta r}\,,
\end{flalign}
\label{eq:added_polar_tractions}
\end{subequations}
where we defined:
\begin{equation}
\overline\Sigma^{(\rm rand)}_{ij}(r,t) =\frac{1}{2\pi}\int_0^{2\pi} \Sigma^{(\rm rand)}_{ij}(r,\theta,t) d\theta\,.
\label{eq:app_angular_average}
\end{equation}
Using the closure for the total stress~\eqref{eq:traction_balance_condition} and the expression for the radial derivative of the velocity profile (Eq.~\eqref{eq:expression_radialderivative_velocity}), the normal condition~\eqref{eq:added_polar_tractions_1} fixes the pressure outside the cavity
\begin{equation}
\mathtt p(R) =-\frac{\Upsilon}{R}-\frac{2\eta\dot R}{R} +\eta_o\left[\partial_r u_\theta(R)-\frac{u_\theta(R)}{R}\right] +\overline\Sigma^{(\rm rand)}_{rr}(R,t)\,.
\label{eq:appendix_normal_traction}
\end{equation}
In particular, a static, nonrotating void would require $\mathtt p(R)=-\Upsilon/R$, consistent with a Laplace pressure jump $p_{\rm inside}-p_{\rm outside}=\Upsilon/R$.
 
Combining the expression for the off-diagonal total stress~\eqref{eq:added_polar_tractions_2} and the tangential traction condition~\eqref{eq:traction_balance_condition}, we obtain another relation:
\begin{equation}
\eta\left[\partial_r u_\theta(R)-\frac{u_\theta(R)}{R}\right] +\frac{2\eta_o\dot R}{R} +\overline\Sigma^{(\rm rand)}_{\theta r}(R,t)=\tau\,.
\label{eq:appendix_tangential_traction}
\end{equation}
The torque density $-\tau$ which generates edge currents is balanced by viscous, odd-viscous, and random shear stresses.

\paragraph{Outer condition and the cutoff $L$.}
An incompressible expanding cavity instantaneously displaces fluid to an outer distance $L$, which can be taken as the size of the box, or a cutoff length for the incompressible regime. The far field pressure is defined here: $\Sigma^{(\rm tot)}_{rr}(L,t)=-p_\infty$.
For a localized vortex, we impose $u_\theta(r)\to0$ at large $r$, and use the infinite-exterior solution when its decay length is small compared with $L-R$. The deterministic far-field pressure then approaches $p_\infty$.

\subsection{Polar equations and the radial force balance}
\label{app:radial_derivation}

The radial and azimuthal components of Eq.~\eqref{eq:appendix_fluctuating_NS} are:
\begin{align}
\rho\left(\partial_t u_r+u_r\partial_r u_r-\frac{u_\theta^2}{r}\right) &=-\gamma\rho u_r-\partial_r\mathtt p +\eta\mathcal L_1u_r+\eta_o\partial_r\omega +f_r^{(\rm rand)}\,,
\label{eq:added_radial_NS}\\
\rho\left(\partial_t u_\theta+\frac{R\dot R}{r}\partial_r u_\theta +\frac{R\dot R}{r^2}u_\theta\right) &=-\gamma\rho u_\theta+\eta\mathcal L_1u_\theta +f_\theta^{(\rm rand)}\,,
\label{eq:appendix_angular_dynamics}
\end{align}
with $\mathcal L_1$ the radial Laplacian and $\omega$ the vorticity within our axisymmetric solution:
\begin{equation}
\mathcal L_1=\partial_r^2+\frac1r\partial_r-\frac1{r^2}\,, \qquad \omega=\partial_r u_\theta+\frac{u_\theta}{r}\,,
\label{eq:app_polar_operators}
\end{equation}
and
\begin{equation}
f_r^{(\rm rand)} =\partial_r\overline\Sigma^{(\rm rand)}_{rr} +\frac{\overline\Sigma^{(\rm rand)}_{rr} -\overline\Sigma^{(\rm rand)}_{\theta\theta}}{r}\,, \qquad f_\theta^{(\rm rand)} =\frac1{r^2}\partial_r \left(r^2\overline\Sigma^{(\rm rand)}_{\theta r}\right)\,.
\label{eq:app_random_force_densities}
\end{equation}

\paragraph{Integrating the radial equation.}
One can simplify Eq.~\eqref{eq:added_radial_NS} by using the expression $u_r=R\dot R/r$ and the expression for the spatial derivatives, such that  $\partial_tu_r=(R\ddot R+\dot R^2)/r$, $u_r\partial_ru_r=-R^2\dot R^2/r^3$, and $\mathcal L_1u_r=0$. 
Integrating Eq.~\eqref{eq:added_radial_NS} in $dr$ from $R$ to $L$ gives
\begin{align}
\rho\left[(R\ddot R+\dot R^2)\log\frac LR -\frac{\dot R^2}{2}\left(1-\frac{R^2}{L^2}\right) -\int_R^L\frac{u_\theta^2}{r} dr\right]=&-\gamma\rho R\dot R\log\frac{L}{R}+\mathtt p(R)-\mathtt p(L)\nonumber\\
& +\eta_o[\omega(L)-\omega(R)] +\int_R^L f_r^{(\rm rand)} dr\,.
\end{align}
The logarithmic dependence follows from the $1/r$ radial flow in two dimensions. A finite system, a collective screening length, or a description allowing density redistribution is needed to regulate it~\cite{marqusee1984dynamics,cates2023classical}.

At the inner boundary, we eliminate $\mathtt p(R)$ using the normal-stress condition, Eq.~\eqref{eq:appendix_normal_traction}. Substituting this expression and using $\omega(R)=\partial_r u_\theta(R)+u_\theta(R)/R$ gives
\begin{equation}
\mathtt p(R)-\eta_o\omega(R)
=-\frac{\Upsilon}{R}-\frac{2\eta\dot R}{R}
-\frac{2\eta_o u_\theta(R)}{R}
+\overline\Sigma^{(\rm rand)}_{rr}(R,t)\,.
\label{eq:app_odd_boundary_cancellation}
\end{equation}
Collecting the boundary contributions and multiplying the integrated radial balance by $2\pi R$ gives, to leading order in $R/L$:
\begin{equation}
\mathcal M(R)\ddot R+\frac12\mathcal M'(R)\dot R^2 +[\mathcal Z_\eta(R)+\mathcal Z_\gamma(R)]\dot R =F[R,u_\theta]+\Xi_\eta(t)\,,
\label{eq:appendix_exact_RP}
\end{equation}
where the coefficients of radial acceleration, substrate drag, and viscous drag are defined by
\begin{equation}
\mathcal M(R)=2\pi\rho R^2\log\frac LR\,, \qquad \mathcal Z_\gamma(R)=\gamma\mathcal M(R)\,, \qquad \mathcal Z_\eta(R)=4\pi\eta\left(1-\frac{R^2}{L^2}\right)\simeq4\pi\eta\,,
\label{eq:appendix_radial_mode}
\end{equation}
the random noise is
\begin{equation}
\Xi_\eta(t)=2\pi R\int_R^L \frac{\overline\Sigma^{(\rm rand)}_{rr}-\overline\Sigma^{(\rm rand)}_{\theta\theta}}{r} dr, \quad\langle\Xi_\eta(t)\Xi_\eta(t')\rangle=16\pi\eta T R^2\int_R^L\frac{dr}{r^3}\delta(t-t')=2T\mathcal Z_\eta(R)\delta(t-t')\,.
\label{eq:added_radial_noise_projection}
\end{equation}
and the instantaneous radial force is
\begin{equation}
F[R,u_\theta]=2\pi R\left[ -p_\infty-\frac{\Upsilon}{R}-\frac{2\eta_o u_\theta(R,t)}{R} +\rho\int_R^L\frac{u_\theta(r,t)^2}{r} dr\right]\,.
\label{eq:appendix_instantaneous_force}
\end{equation}
This force retains the full time dependence of $u_\theta$.

Eqs.~\eqref{eq:appendix_exact_RP} and~\eqref{eq:appendix_angular_dynamics}, together with the tangential condition~\eqref{eq:appendix_tangential_traction}, form a coupled system of equations for $R$ and $u_\theta$. To obtain an evolution equation for $R$ alone, we now need to eliminate $u_\theta$. 

\subsection{Stationary vortex and mechanical work}

We first consider for simplicity a stationary cavity without random forces. For such a case, $u_r$, $\dot R$, $\ddot R$ and $\partial_tu_\theta$ all vanish. With $\ell_\gamma=\sqrt{{\eta}/{\gamma\rho}}$, the solution to the azimuthal velocity can be obtained by setting the right-hand side of Eq.~\eqref{eq:appendix_angular_dynamics} equal to zero (with $f_\theta^{(\rm rand)}=0$). The solution has the following expression:
\begin{equation}
u_\theta^{\rm st}(r;R)=V_R\frac{K_1(r/\ell_\gamma)}{K_1(R/\ell_\gamma)}=-\dfrac{\tau \ell_\gamma}{\eta}\dfrac{K_1(r/\ell_\gamma)}{K_2(R/\ell_\gamma)}\,,  \qquad V_R=-\dfrac{\tau\ell_\gamma} {\eta}\dfrac{K_1(R/\ell_\gamma)}{K_2(R/\ell_\gamma)}\,,
\label{eq:appendix_stationary_vortex}
\end{equation}
with $K_n$ the $n$th modified Bessel function of the second kind.
Substitution into Eq.~\eqref{eq:appendix_instantaneous_force} gives
\begin{equation}
F(R)\equiv F[R, u_\theta^{\rm st}(\,\cdot\,; R)]=2\pi R\left(\frac{\rho V_R^2}{2}\left[1-\left(\frac{K_0(R/\ell_\gamma)}{K_1(R/\ell_\gamma)}\right)^2\right]-p_\infty-\frac{\Upsilon}{R}-\frac{2\eta_oV_R}{R}\right)\,.
\label{eq:appendix_stationary_force}
\end{equation}
This recovers the result presented in the main text, which describes stationary states and bubble dynamics in the limit of instantaneous flow relaxation. To account for finite relaxation times in the flow and nucleation, we return to the coupled equations and retain corrections from a nonzero $\dot R$.

\subsection{Perturbative relaxation of the vortex}
\label{app:vortex_lag}

From now on, we set $\eta_o=0$ to streamline the derivation. This assumption is not necessary, but simplifies the analysis and corresponds to the case used for comparison with the simulation.

We now retain the full time dependence rather than restricting to the stationary state. Since Eq.~\eqref{eq:appendix_exact_RP} depends on $u_\theta(t)$ through $F[R,u_\theta(t)]$, closing it requires solving Eq.~\eqref{eq:appendix_angular_dynamics} for $u_\theta(t)$ in terms of $R(t)$, at least perturbatively.
Writing $u_\theta(r,t)=u_\theta^{\rm st}(r;R(t))+\delta u_\theta(r,t)$, with $u_\theta^{\rm st}$ from Eq.~\eqref{eq:appendix_stationary_vortex} and using the stationary equation from Eq.~\eqref{eq:appendix_angular_dynamics} gives
\begin{equation}
\partial_t\delta u_\theta+\gamma\mathcal B\delta u_\theta =-\frac\tau\eta s\dot R+\frac{f_\theta^{(\rm rand)}}\rho -\frac{R\dot R}{r}\left(\partial_r\delta u_\theta+\frac{\delta u_\theta}{r}\right)\,.
\label{eq:added_untruncated_angular_response}
\end{equation}
Here $\mathcal B$ describes vortex relaxation and $s$ is the source generated by cavity motion. In terms of the dimensionless quantities
\begin{equation}
\begin{gathered}
\widetilde u_\theta^{\rm st}(x;a)=\frac{\eta}{\tau\ell_\gamma}u_\theta^{\rm st}(r;R)=-\frac{K_1(x)}{K_2(a)}\,,\qquad x=\frac r{\ell_\gamma}\,,\qquad a=\frac R{\ell_\gamma}\,,\qquad \lambda_\tau=\frac{\tau}{\eta\gamma}\,,
\end{gathered}
\label{eq:appendix_dimensionless_vortex}
\end{equation}
the operator $\mathcal B$ and source $s$ are
\begin{equation}
\mathcal B=-\partial_x^2-\frac1x\partial_x+1+\frac1{x^2}\,, \qquad s(x;a)=\partial_a\widetilde u_\theta^{\rm st}+\frac a x\partial_x\widetilde u_\theta^{\rm st}+\frac a{x^2}\widetilde u_\theta^{\rm st}=\frac{K_1(x)K_2'(a)}{K_2(a)^2} +\frac{aK_0(x)}{xK_2(a)}\,.
\label{eq:app_response_operator_source}
\end{equation}
The source $s$ comes from the radial advection of the stationary profile evaluated at $R(t)$. By the chain rule:
\begin{equation}
\left.\partial_tu_\theta^{\rm st}\right|_r+\frac{R\dot R}{r}\left(\partial_ru_\theta^{\rm st}+\frac{u_\theta^{\rm st}}{r}\right)=\frac\tau\eta\dot Rs(x;a)\,.
\label{eq:added_moving_base_derivatives}
\end{equation}
The last term in Eq.~\eqref{eq:added_untruncated_angular_response} is of order $\dot R \delta u_\theta$ and is negligible in a joint first-order expansion in the boundary speed and velocity perturbation. Hence, we obtain
\begin{equation}
\partial_t\delta u_\theta+\gamma\mathcal B\delta u_\theta =-\frac\tau\eta s\dot R+\frac{f_\theta^{(\rm rand)}}\rho\,.
\label{eq:appendix_linear_angular_response}
\end{equation}
The boundary condition for the perturbation does not include the torque density, which is already accounted for in the stationary profile. Hence, Eq.~\eqref{eq:appendix_tangential_traction} gives
\begin{equation}
\eta\left(\partial_r-\frac{1}{R}\right)\delta u_\theta(R,t) =-\overline\Sigma^{(\rm rand)}_{\theta r}(R,t)\,.
\label{eq:added_perturbed_vortex_traction}
\end{equation}

\subsection{Slow vortex response}
\label{app:slow_vortex_response}

For slow radius dynamics, eliminating the velocity correction gives a contribution to radial friction and a random radial force:
\begin{equation}
F[R,u_\theta]\simeq F(R)-\mathcal Z_\theta(R)\dot R+\Xi_\theta(t)\,.
\label{eq:appendix_eliminated_vortex_force}
\end{equation}
We first obtain the common velocity response, then calculate $\mathcal Z_\theta$ from the mean vortex lag and $\Xi_\theta$ from the vortex fluctuations.

When the radius and its speed vary slowly compared with vortex relaxation, we neglect $\partial_t\delta u_\theta$ in Eq.~\eqref{eq:appendix_linear_angular_response}. The time dependence of $u_\theta^{\rm st}(r;R(t))$ remains included in $s$. The response equation becomes
\begin{equation}
\gamma\mathcal B\delta u_\theta=-\frac\tau\eta s\dot R+\frac{f_\theta^{(\rm rand)}}\rho\,,
\label{eq:app_slow_angular_response}
\end{equation}
with the random boundary traction in Eq.~\eqref{eq:added_perturbed_vortex_traction}. This approximation requires
\begin{equation}
\omega_R\tau_\theta\ll1\,, \qquad \tau_\theta^{-1}\sim\gamma+\frac{\eta}{\rho R^2}\,,
\label{eq:appendix_angular_adiabaticity}
\end{equation}
with $\omega_R$ the characteristic rate of radius evolution. We therefore obtain from Eq.~\eqref{eq:app_slow_angular_response}:
\begin{equation}
\begin{gathered}
u_\theta(r,t)\simeq u_\theta^{\rm st}(r;R(t))+\delta u_\theta^{(\rm det)}(r,t)+\delta u_\theta^{(\rm rand)}(r,t)\,,\\
\delta u_\theta^{(\rm det)}(r,t)=-\lambda_\tau\dot R\psi(x;a)\,, \qquad
\psi(x;a)=\int_a^\infty yG_a(x,y)s(y;a)dy\,,\\
\delta u_\theta^{(\rm rand)}(r,t)=-\frac{\ell_\gamma}{\eta}\int_a^\infty \left[y\partial_yG_a(x,y)-G_a(x,y)\right]\overline\Sigma^{(\rm rand)}_{\theta r}(\ell_\gamma y,t)dy\,.
\end{gathered}
\label{eq:app_slow_vortex_solution}
\end{equation}
Here $G_a$ is the Green function of $\mathcal B$ with homogeneous shear condition at $x=a$ and decay at infinity:
\begin{equation}
G_a(x,y)=I_1(\min\{x,y\})K_1(\max\{x,y\}) +\frac{I_2(a)}{K_2(a)}K_1(x)K_1(y)\,, \qquad \mathcal B_xG_a(x,y)=\frac{\delta(x-y)}y\,.
\label{eq:added_vortex_Green_function}
\end{equation}
Thus $\psi$ is the response to the known source $s$. 
Inserting $\delta u_\theta = \delta u_\theta^{(\rm det)} + \delta u_\theta^{(\rm rand)}$ into Eq.~\eqref{eq:appendix_instantaneous_force} yields two contributions defined in Eq.~\eqref{eq:appendix_eliminated_vortex_force}:
\begin{equation}
\mathcal Z_\theta(R)=2\pi\eta\lambda_\tau^2 I_Z(a)\,, \qquad I_Z(a)=2a\int_a^\infty\frac{\widetilde u_\theta^{\rm st}(x;a)\psi(x;a)}x dx\, ,
\label{eq:appendix_Ztheta}
\end{equation}
and
\begin{equation}
\Xi_\theta(t)=-4\pi\lambda_\tau R\int_a^\infty x\left(\varphi'-\frac\varphi x\right)\overline\Sigma^{(\rm rand)}_{\theta r}(\ell_\gamma x,t) dx\,, \qquad \varphi(x;a)=\int_a^\infty G_a(x,y)\frac{\widetilde u_\theta^{\rm st}(y;a)}y dy\,.
\label{eq:added_explicit_tangential_noise}
\end{equation}
Exchanging the two integrals defines the weight $\varphi$.
The resulting noise strength is
\begin{align}
\langle\Xi_\theta(t)\Xi_\theta(t')\rangle &=2\mathcal A_\theta(R)\delta(t-t')\,,\quad
\frac{\mathcal A_\theta(R)}T =2\pi\eta\lambda_\tau^2I_A(a)\,, \quad I_A(a)=4a^2\int_a^\infty x\left(\varphi'-\frac\varphi x\right)^2dx\,.
\label{eq:appendix_Atheta}
\end{align}

\subsection{Overdamped radius dynamics and the nucleation exponent}
\label{app:scalar_nucleation}

We now neglect the inertial terms in Eq.~\eqref{eq:appendix_exact_RP} and substitute the vortex force from Eq.~\eqref{eq:appendix_eliminated_vortex_force}. Using $F(R)=-W'(R)$ and adding the noise covariances in Eqs.~\eqref{eq:added_radial_noise_projection} and~\eqref{eq:appendix_Atheta}, we obtain the overdamped radius dynamics:
\begin{equation}
\mathcal Z(R)\dot R=-W'(R)+\Xi_R(t)\,, \qquad \langle\Xi_R(t)\Xi_R(t')\rangle =2T H(R)\delta(t-t')\,.
\label{eq:appendix_reduced_radius}
\end{equation}
where
\begin{equation}
\mathcal Z(R)=\mathcal Z_\eta+\mathcal Z_\gamma+\mathcal Z_\theta\,, \qquad H(R)=\mathcal Z_\eta+\frac{\mathcal A_\theta}{T}\,, \qquad \Xi_R(t)=\Xi_\eta(t)+\Xi_\theta(t)\,.
\label{eq:app_total_radial_coefficients}
\end{equation}
Neglecting inertia requires $\mathcal Z>0$, $\omega_R\mathcal M/\mathcal Z\ll1$, and a negligible quadratic inertial term. 

\paragraph{Neglecting the external damping drag}
Within our incompressible description, the drag related to external damping diverges with system size:
\begin{equation}
\frac{\mathcal Z_\gamma}{\mathcal Z_\eta} \simeq\frac12\left(\frac R{\ell_\gamma}\right)^2\log\frac{L}{R}\,.
\label{eq:added_substrate_drag_ratio}
\end{equation}
In a realistic system, $L$ would instead correspond to a sound propagation cutoff length. Since we have weak damping, relatively low density, and strong chiral edge currents within our system, we use the approximation $\mathcal Z_\gamma\ll\mathcal Z_\eta+\mathcal Z_\theta$ over the relevant radii. In that regime $\mathcal Z\simeq\mathcal Z_\eta+\mathcal Z_\theta$, while $H$ is unchanged. 

\paragraph{Weak-noise quasipotential.}
Dividing Eq.~\eqref{eq:appendix_reduced_radius} by $\mathcal Z$ gives the deterministic drift $b=-W'/\mathcal Z$ and diffusion coefficient $D_R=TH/\mathcal Z^2$. In an It\^o representation, the probability current has the form
\begin{equation}
J=\left[-\frac{W'}{\mathcal Z}+\mathcal O(T)\right]P -\partial_R\left(\frac{TH}{\mathcal Z^2}P\right)\,,
\label{eq:added_radius_probability_current}
\end{equation}
with $P(R)$ the (quasi-)probability of observing a radius $R$. The unspecified order-$T$ term includes noise-induced drift from the fast-variable elimination or a change of convention, which do not change the leading weak-noise large deviation barrier~\cite{cates2023classical}.
By expressing 
\begin{equation}
P(R)\asymp\exp[-\Phi(R)/T]\,,
\label{eq:appendix_probability_definition}
\end{equation}
the leading zero-current balance inside a metastable basin is
\begin{equation}
0=-\frac{W'}{\mathcal Z}+\frac H{\mathcal Z^2}\Phi'+\mathcal O(T)\Rightarrow \Phi'(R)=W'(R)\frac{\mathcal Z(R)}{H(R)}\,.
\label{eq:added_quasipotential_derivation}
\end{equation}

Using $\mathcal Z_\eta\simeq4\pi\eta$ and neglecting $\mathcal Z_\gamma$ under the criterion above yields the large deviation barrier:
\begin{equation}
\Phi_c = \int_{0}^{R_c}W'(R)\frac{\mathcal Z(R)}{H(R)} dR=\int_0^{R_c}W'(R) \frac{1+\lambda_\tau^2 I_Z(R/\ell_\gamma)/2} {1+\lambda_\tau^2 I_A(R/\ell_\gamma)/2} dR\,.
\label{eq:appendix_nucleation_barrier}
\end{equation}

For $q=\mathcal Z/H$, the barriers satisfy
\begin{equation}
q_{\min}  W_c\le\Phi_c\le q_{\max} W_c\,.
\label{eq:appendix_barrier_bound}
\end{equation}
Thus, a bounded $q$ makes the two barriers comparable. As an example, at $\phi=0.55$, $\alpha=0.1$, and $\gamma\sigma/|\Delta|=0.05$, this reduced calculation gives $\Phi(R_c)/W(R_c)\simeq3$. 

Under this closure, $W(R_c)$ sets the barrier scale, although not its exact value. A more complete treatment would retain radial inertia and account for the memory and colored noise associated with finite vortex relaxation. Direct computation of the reaction path using a deep neural network could also be done~\cite{zakine2024unveiling}.

\section{Limiting cases of radial stability}
\label{app:stability} 

We set $\Upsilon=0$, and distinguish the radial traction $f(R)\equiv F(R)/(2\pi R)$ from the total force $F=-W'$.

\subsection{Undamped fluid}
\label{app:undamped}

\begin{figure}
    \centering
    \includegraphics[width=0.9\linewidth]{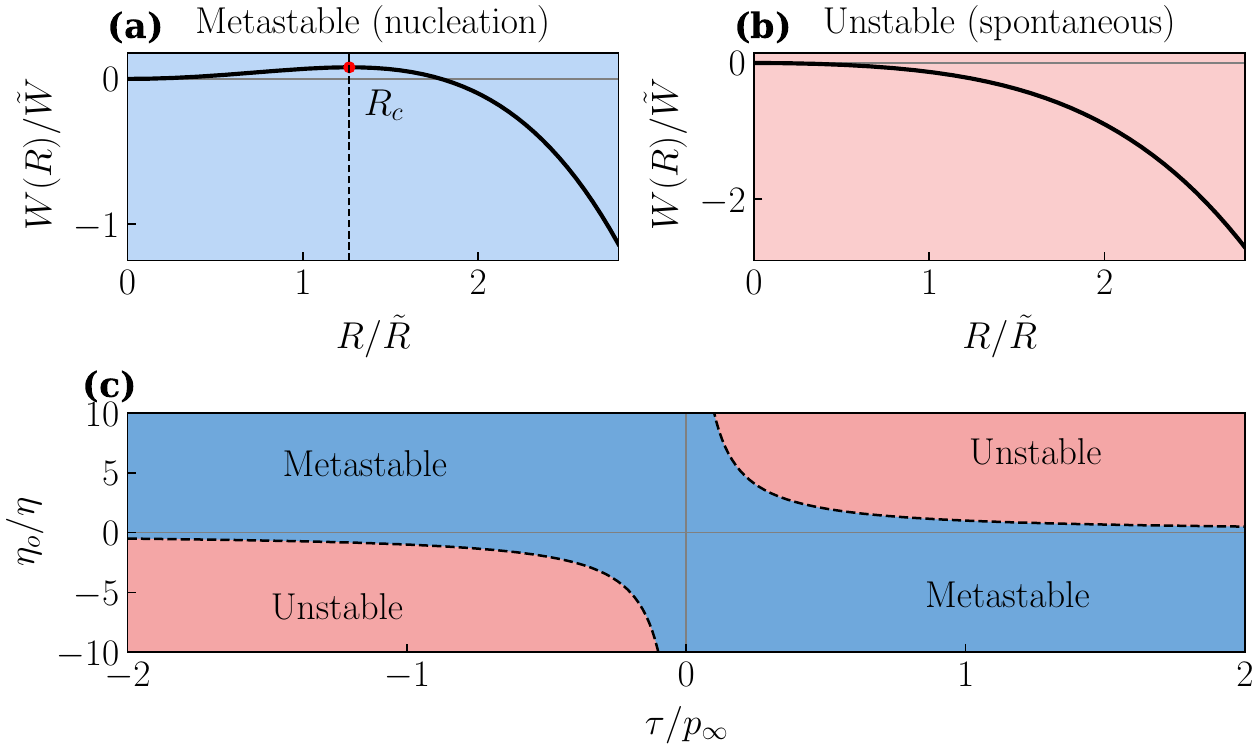}
    \caption{Undamped radial dynamics with $\gamma=\Upsilon=0$.
    (a) A metastable homogeneous state, with work barrier at $R_c$.
    (b) An unstable homogeneous state, with barrierless growth.
    (c) Phase diagram; the boundary is $\eta p_\infty=\eta_o\tau$, as given by Eq.~\eqref{eq:stability_criterion}.
    Panels (a,b) use $p_\infty/|\tau|-|\eta_o|/\eta=\{0.2,-0.2\}$, respectively, with $|\eta_o|/\eta=1$ and $\tau\eta_o>0$. Panel (c) shows both signs of $\tau/p_\infty$ and $\eta_o/\eta$. Opposite signs give metastability throughout the undamped theory. Units are $\tilde R=\eta/\sqrt{\rho|\tau|}$ and $\tilde W=2\pi|\tau|\tilde R^2$.}
    \label{fig:nodamping}
\end{figure}

Taking the vanishing-damping limit, $\gamma\to0$ (and hence $\ell_\gamma\to\infty$), the net radial force $F(R)$ given by Eq.~\eqref{eq:stationary_void_force}, which governs the expansion or collapse of the bubble, reduces to
\begin{equation}
    \frac{F(R)}{2\pi R}=-p_\infty+\tau \dfrac{\eta_o}{\eta}+\rho  \dfrac{\tau^2R^2}{8\eta^2}\,.
    \label{eq:force_no_damping}
\end{equation}
Pressure closes the cavity while the constant shift $\tau\eta_o/\eta$ favors opening or closure according to its sign. The centrifugal contribution per unit circumference grows as $R^2$.

By setting $F(R)=0$ (corresponding to $W'(R)=0$) and solving for $R>0$, we look for the existence of a critical radius $R_c$ and a stable radius $R_b$. 
Equation~\eqref{eq:force_no_damping} admits a non-vanishing root in the regime of low torque density or odd viscosity, when the following condition is satisfied: $\eta p_\infty-\eta_o\tau>0$, with the following expression 
\begin{equation}
    R_c=\sqrt{\dfrac{8\eta }{\rho\tau^2}}\sqrt{\eta p_\infty - \eta_o\tau}\,.
    \label{eq:compatible_linear_stability}
\end{equation}
This solution can be identified with $R_c$ since it corresponds to a maximum of $W$, as shown in Fig.~\ref{fig:nodamping}(a).
Cavities smaller than $R_c$ collapse because the centrifugal term is negligible, while cavities larger than $R_c$ grow. The homogeneous state is therefore metastable. With no damping, $W$ is unbounded below, and growth continues endlessly until a finite system size eventually limits it. Shape instability may intervene before then (Appendix~\ref{app:noncircular}).
If instead
\begin{equation}
    \eta p_\infty - \eta_o\tau<0\,,
    \label{eq:stability_criterion}
\end{equation}
then $f(0)>0$, $R_c$ is not real, and arbitrarily small cavities grow, as shown in Fig.~\ref{fig:nodamping}(b). Odd viscosity and torque overcome pressure even without centrifugal forcing; inertia accelerates subsequent growth. Eq.~\eqref{eq:stability_criterion} is the nonlinear counterpart of the homogeneous linear-instability threshold in Ref.~\onlinecite{Marconi2026hydrodynamics}. The odd-viscous shift acts like an effective vapor pressure in cavitation~\cite{van2016dynamics}. The phase boundary in Fig.~\ref{fig:nodamping}(c) is given by $\eta p_\infty=\eta_o\tau$.

Small $R_c$ and $W(R_c)$ make rapid nucleation difficult to distinguish from instability; a large barrier can make metastability appear stable over accessible times. Line tension adds a closing traction at small $R$, but its continuum description fails near a particle diameter. A small tension therefore need not regularize microscopic instability, consistent with the Korteweg-stress analysis of Ref.~\onlinecite{Marconi2026hydrodynamics}.

\subsection{Damped fluid without odd viscosity}
\label{app:no_odd_viscosity}

\begin{figure}
    \centering
    \includegraphics[width=0.9\linewidth]{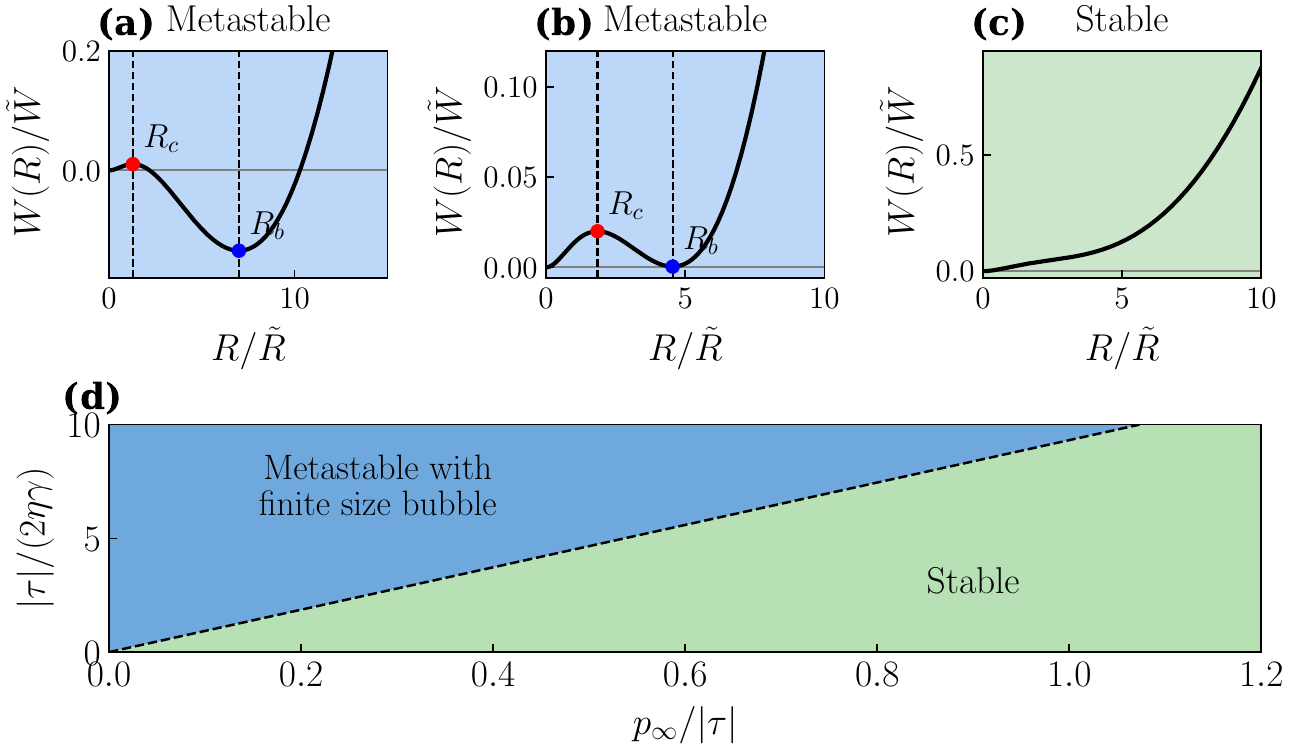}
    \caption{Damped radial dynamics without odd viscosity ($\eta_o=0$).
    (a,b) A metastable homogeneous state, separated from a finite-radius bubble at $R_b$ by the work barrier at $R_c$.
    (c) A stable homogeneous state with no finite-radius minimum.
    (d) Phase diagram, with boundary given by equality in Eq.~\eqref{eq:separatrix}.
    Panels (a--c) use $p_\infty/|\tau|=\{0.042,0.05,0.06\}$ and $|\tau|/(\gamma\eta)=1$; (d) varies $p_\infty/|\tau|$ and $|\tau|/(2\eta\gamma)$. Units are $\tilde R=\eta/\sqrt{\rho|\tau|}$ and $\tilde W=2\pi|\tau|\tilde R^2$.}
    \label{fig:noeta}
\end{figure}

For $\gamma>0$ and $\eta_o=0$, expansion is driven solely by the centrifugal effect of the torque-induced vortex. Eq.~\eqref{eq:stationary_void_force} becomes
\begin{equation}
    f(R)= -p_\infty+\frac{\tau^2}{2\eta\gamma} g (x)\,, 
    \qquad g(x)=\frac{K_1(x)^2-K_0(x)^2}{K_2(x)^2}\,,  \qquad x=\frac{R}{\ell_\gamma}\,.
    \label{eq:force_without_odd_visco}
\end{equation}
Both $f(0)$ and $f(\infty)$ equal $-p_\infty$, hence both small and very large cavities shrink. A finite bubble exists if $f(R)>0$ for a range of radii. Since $g$ has a single maximum $g_{\max}\simeq0.107$ at $x_{\max}\simeq2.86$, two distinct roots exist precisely when
\begin{equation}
    p_\infty < \frac{\tau^2}{2\eta\gamma} g_{\max}
     \quad \Rightarrow 
    \quad \eta\gamma p_\infty \lesssim 0.053 \tau^2\,.
    \label{eq:separatrix}
\end{equation}
At threshold, $R_c=R_b=x_{\max}\ell_\gamma$. Deep in the bubble regime, $p_\infty\ll g_{\max}\tau^2/(2\eta\gamma)$, the asymptotic forms $g(x\to0)\sim x^2/4$ and $g(x\to\infty)\sim1/x$ give
\begin{equation}
    R_c \simeq \sqrt{\frac{8\eta^2 p_\infty}{\rho \tau^2}}\,, 
    \qquad R_b \simeq \frac{\tau^2}{2p_\infty\sqrt{\rho \eta \gamma^3}} \,.
\end{equation}
Thus $R_b$ grows with torque and decreases with pressure and damping, while $R_c$ marks the smallest growing seed.

Figures~\ref{fig:noeta}(a) and (b) show the resulting maximum at $R_c$ and radially stable minimum at $R_b$. Because $\eta_o=0$, both cases require nucleation. We note that a switching between the two minima may account for repeated bubble appearance and disappearance in Ref.~\onlinecite{shen2023collective}.

When dissipation overwhelms centrifugal forcing, no finite-radius minimum exists, as shown in Fig.~\ref{fig:noeta}(c). Panel (d) summarizes this stable regime and the metastable regime. The stable region was absent from the nonlinear analysis of Ref.~\onlinecite{Marconi2026hydrodynamics}, which set $\eta=0$. Finite shear viscosity regularizes the exterior flow and permits stable large-scale dynamics here.

\section{Stability to noncircular perturbations}
\label{app:noncircular}
\label{sec:noncircular}

\begin{figure}
    \centering
    \includegraphics[width=0.99\linewidth]{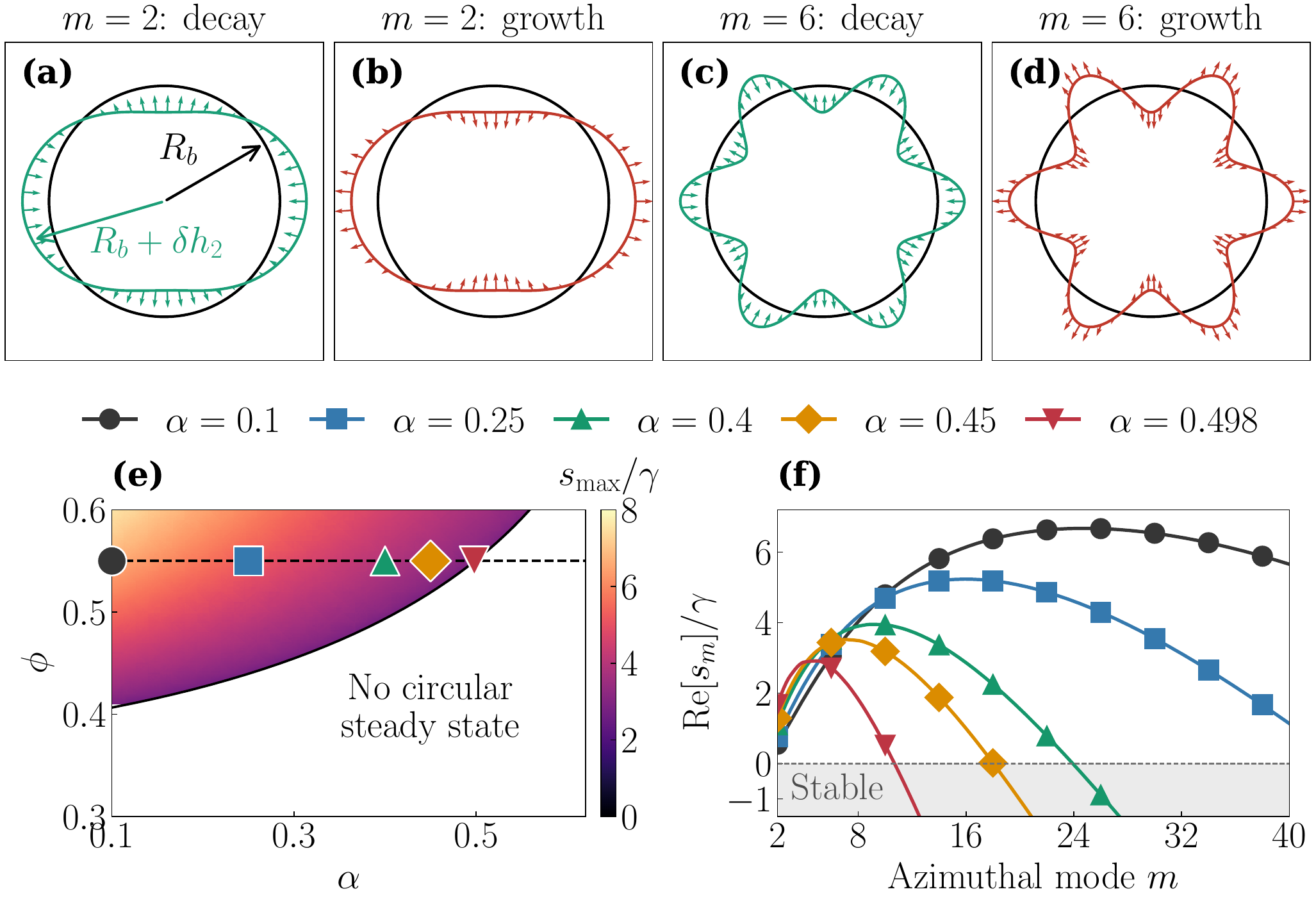}
    \caption{Shape stability of the stationary bubble.
    (a--d) Schematic decay and growth of $m=2$ and $m=6$ perturbations about the black circle.
    (e) Largest growth rate $s_{\max}$ at $R_b$, using $\eta_o=0$, and $\gamma\sigma/|\Delta|=0.05$. Colors show unit-width intervals of $s_{\max}/\gamma$. White regions have no stationary bubble.
    (f) Spectra along the dashed cut $\phi=0.55$, at the marked state points.}
    \label{fig:instability_figure}
\end{figure}

We test whether a stationary bubble of radius $R=R_b$ remains circular. The unperturbed flow is $V(r)\bm e_\theta$, given by Eq.~\eqref{eq:stationary_void_vortex}. We perturb the interface and surrounding flow with an angular mode $m\ge2$,
\begin{equation}
r_s(\theta,t)=R+h e^{st+im\theta}\,,\qquad (\delta u_r,\delta u_\theta,\delta P)=(u(r),v(r),\widehat P(r))e^{st+im\theta}\,.
\label{eq:noncircular_ansatz}
\end{equation}
Here $h$ is the interface-displacement amplitude, $s$ is the complex growth rate, and $u$, $v$, and $\widehat P$ are radial profiles. We use the reduced pressure $P=\mathtt p-\eta_o\omega$, where $\omega=(\bm\nabla\times\bm u)_z$, to absorb odd viscosity into the bulk pressure. The end goal will be to see whether an infinitesimal perturbation of a mode $m$ grows or shrinks under time evolution, as shown in Fig.~\ref{fig:instability_figure}(a-d). 
Taking the curl of Eq.~\eqref{eq:odd-navier_stokes} gives
\begin{equation}
\partial_t\omega+\bm u\cdot\bm\nabla\omega=-\gamma\omega+\nu\bm\nabla^2\omega\,,\qquad \nu={\eta}/{\rho}\,.
\label{eq:noncircular_vorticity_dynamics}
\end{equation}
The base flow has vorticity $V'+V/r$. Writing its perturbation as $w(r)e^{st+im\theta}$ and subtracting the stationary equation yields
\begin{equation}
\left(s+\gamma+\frac{imV}{r}\right)w+u\left(V'+\frac Vr\right)'=\nu\left(w''+\frac{w'}r-\frac{m^2}{r^2}w\right)\,.
\label{eq:noncircular_linear_vorticity}
\end{equation}
With $u=im\psi/r$, $v=-\psi'$, and $w=-L_m\psi$, where $L_m=\partial_r^2+r^{-1}\partial_r-m^2/r^2$ is the radial Laplacian operator for the $m$-th mode. The linearized vorticity equation~\eqref{eq:noncircular_linear_vorticity} becomes:
\begin{equation}
\nu L_m^2\psi-\left(s+\gamma+\frac{imV}{r}\right)L_m\psi+\frac{im}{r}\left(V'+\frac Vr\right)'\psi=0\,,\qquad r>R\,.
\label{eq:noncircular_eigenproblem}
\end{equation}
All coefficients are fixed by the stationary profile $V(r)$, for which $(V'+V/r)'=V/\ell_\gamma^2$.

At the moving interface, three conditions are required. The boundary moves with the fluid, and both tangential and normal tractions vanish. Linearizing these conditions at the displaced and tilted boundary gives, respectively,
\begin{subequations}
\label{eq:noncircular_boundary}
\begin{align}
u_R&=(s+imV_R/R)h\,,\label{eq:noncircular_kinematic_bc}\\
0&=\eta\left(v'-\frac vR+\frac{im u}{R}\right)-2\eta_o u'
+\left(\rho\gamma V_R-\frac{2\tau}{R}+\frac{2im\eta_o\tau}{\eta R}\right)h\,,\label{eq:noncircular_tangential_bc}\\
0&=-\widehat P+2\eta u'+\frac{2\eta_o}{R}(im u-v)
-\left(\frac{\rho V_R^2}{R}+\frac{2\eta_o\tau}{\eta R}+\frac{2im\tau}{R}\right)h\,.\label{eq:noncircular_normal_bc}
\end{align}
\end{subequations}
All quantities in these conditions are evaluated at $R^+$, with $u_R=u(R^+)$ and $V_R=V(R^+)$. The terms proportional to $h$ account for displacement and tilting of the interface.  The pressure can be obtained from the azimuthal velocity field equation:
\begin{equation}
\rho\left[\left(s+\gamma+\frac{imV}{r}\right)v+\left(V'+\frac Vr\right)u\right]=-\frac{im}{r}\widehat P+\eta\left[\left(L_m-\frac{1}{r^2}\right)v+\frac{2im}{r^2}u\right]\,.
\label{eq:noncircular_azimuthal_momentum}
\end{equation}
Using the streamfunction yields
\begin{equation}
\widehat P=\frac{r}{im}\left[\rho\left(s+\gamma+\frac{imV}{r}\right)\psi'-\eta(L_m\psi)'\right]-\rho\left(V'+\frac Vr\right)\psi\,.
\label{eq:noncircular_pressure}
\end{equation}
Decay at infinity leaves two independent solutions of the fourth-order equation for $\psi$. A nonzero solution exists only for particular growth rates $s$, which define the eigenvalue problem. The overall perturbation amplitude remains arbitrary. Modes $m\ge2$ preserve the cavity area to first order and therefore test shape stability independently of radial stability. If $s_m$ is the eigenvalue with the largest real part for mode $m$, the circle is unstable when
\begin{equation}
\mathrm{Re}[s_m]>0\text{ for some }m\geq 2\,.
\label{eq:instability_condition}
\end{equation}
Figure~\ref{fig:instability_figure}(e,f) shows growing shape modes throughout the sampled nucleation region at $R_b$, so radial stability generally does not preserve a circle. Along $\phi=0.55$, increasing $\alpha$ from $0.10$ to $0.498$ lowers $s_{\max}/\gamma$ from about $6.7$ to $2.9$ and shifts the fastest mode from $m=25$ to $m=5$. Growth remains positive near the branch endpoint, and the dominant deformation need not be elliptical. 

Growing modes could saturate into an ellipse or another persistent shape. Faster growth deeper in the nucleation region could instead favor large deformations and splitting. A nonlinear calculation is required to settle this matter, as the eigenproblem alone predicts neither the final shape nor breakup.

\printbibliography

\end{document}